\documentclass[journal,twoside]{IEEEtran}
\pdfoutput=1
\UseRawInputEncoding
\usepackage{amsfonts}
\usepackage{cite}
\usepackage{amsmath}
\usepackage{extarrows}
\usepackage{bm}
\usepackage{booktabs}
\usepackage{xcolor}
\usepackage{algorithmic}
\usepackage{textcomp}
\usepackage{stfloats}
\usepackage{enumerate}
\usepackage{footnote}
\ifCLASSINFOpdf
   \usepackage[pdftex]{graphicx}
   \graphicspath{{../pdf/}{../jpeg/}}
   \DeclareGraphicsExtensions{.pdf,.jpeg,.png}
\else
   \usepackage[dvips]{graphicx}
   \graphicspath{{../eps/}}
   \DeclareGraphicsExtensions{.eps}
\fi
\begin{document}
\title{Plane Spiral OAM Mode-Group Based MIMO Communications: An Experimental Study}
\author{Xiaowen~Xiong,
        Shilie~Zheng,
        Zelin~Zhu,
        Yuqi~Chen,
        Hongzhe~Shi,
        Bingchen~Pan,
        Cheng~Ren,
        Xianbin~Yu,
        Xiaofeng~Jin,
        Wei E.I. Sha,
        and~Xianmin~Zhang% <-this % stops a space
\thanks{X. Xiong, S. Zheng, Z. Zhu, Y. Chen, B. Pan, C. Ren, X. Jin and W. E. I. Sha are with the College of Information Science and Electronic Engineering, Zhejiang University, Hangzhou 310027, China (e-mail: zhengsl@zju.edu.cn). H. Shi is with the Huawei Technologies Ltd., Shanghai 210206, China. X. Yu is with the Zhejiang Lab, Hangzhou 310000, China, and is also with the College of Information Science and Electronic Engineering, Zhejiang University, Hangzhou 310027, China. 
X. Zhang is with the Ningbo Research Institute, Zhejiang University, Ningbo 315100, and is also with the College of Information Science and Electronic Engineering, Zhejiang University, Hangzhou 310027, China (e-mail: zhangxm@zju.edu.cn).}% <-this % stops a space
}

\markboth{}%
{Xiong \MakeLowercase{\textit{et al.}}}

\maketitle

%\vspace*{-12mm}
\begin{abstract}
Spatial division multiplexing using conventional orbital angular momentum (OAM) has become a well-known physical layer transmission method over the past decade. The mode-group (MG) superposed by specific single mode plane spiral OAM (PSOAM) waves has been proved to be a flexible beamforming method to achieve the azimuthal pattern diversity, which inherits the spiral phase distribution of conventional OAM wave. Thus, it possesses both the beam directionality and vorticity. In this paper, it's the first time to show and verify novel PSOAM MG based multiple-in-multiple-out (MIMO) communication link (MG-MIMO) experimentally in a line-of-sight (LoS) scenario. A compact multi-mode PSOAM antenna is demonstrated experimentally to generate multiple independent controllable PSOAM waves, which can be used for constructing MGs. After several proof-of-principle tests, it has been verified that the beam directionality gain of MG can improve the receiving signal-to-noise (SNR) level in an actual system, meanwhile, the vorticity can provide another degree of freedom (DoF) to reduce the spatial correlation of MIMO system. Furthermore, a tentative long-distance transmission experiment operated at 10.2 GHz has been performed successfully at a distance of 50 m with a single-way spectrum efficiency of 3.7 bits/s/Hz/stream. The proposed MG-MIMO may have potential in the long-distance LoS back-haul scenario.
\end{abstract}

% \fi
\IEEEpeerreviewmaketitle

\section{Introduction}
\IEEEPARstart{T}{he} capacity demand of wireless communication has been growing rapidly in the past decade, 1 Tera-bits/s throughput has been expected to achieve in the next generation wireless communication networks\cite{6g}. From the perspective of physical layer, various orthogonal resources have been investigated to improve the capacity performance, such as time, frequency and space\cite{wang20206g}. As a distinct space division multiplexing technique, radio frequency (RF) orbital angular momentum (OAM) waves have been considered as one type of spatial orthogonal basis sets to achieve same-frequency mode division multiplexing (MDM) \cite{mohammadi2009orbital}, in which OAM waves present spiral phase front of $e^{-jl\varphi}$ ($l$ is OAM mode order) \cite{allen1992orbital}. 

Since the famous \emph{Venice experiment}\cite{tamburini2012encoding} was reported, a series of OAM based MDM experiments in the line-of-sight (LoS) scenario have been published continually\cite{yan2014high,hui2015multiplexed,zhang2016mode,sasaki2020experimental}. In \cite{sasaki2020experimental}, NTT Corporation has demonstrated an exhilarating work, in which they have successfully carried out an experiment of 100 Gbit/s OAM transmission at a distance of 100 m. Besides, OAM wave has been extended to other application fields including radar imaging and detection\cite{chen2017single,liu2019microwave}, eigenmode-beamforming\cite{zheng2017realization,zhu2020compact} and rotational doppler measurement\cite{zheng2018analysis}, etc. 
\begin{figure}[t]
%    \vspace*{-8mm}
	\centering
	\includegraphics[width=3in]{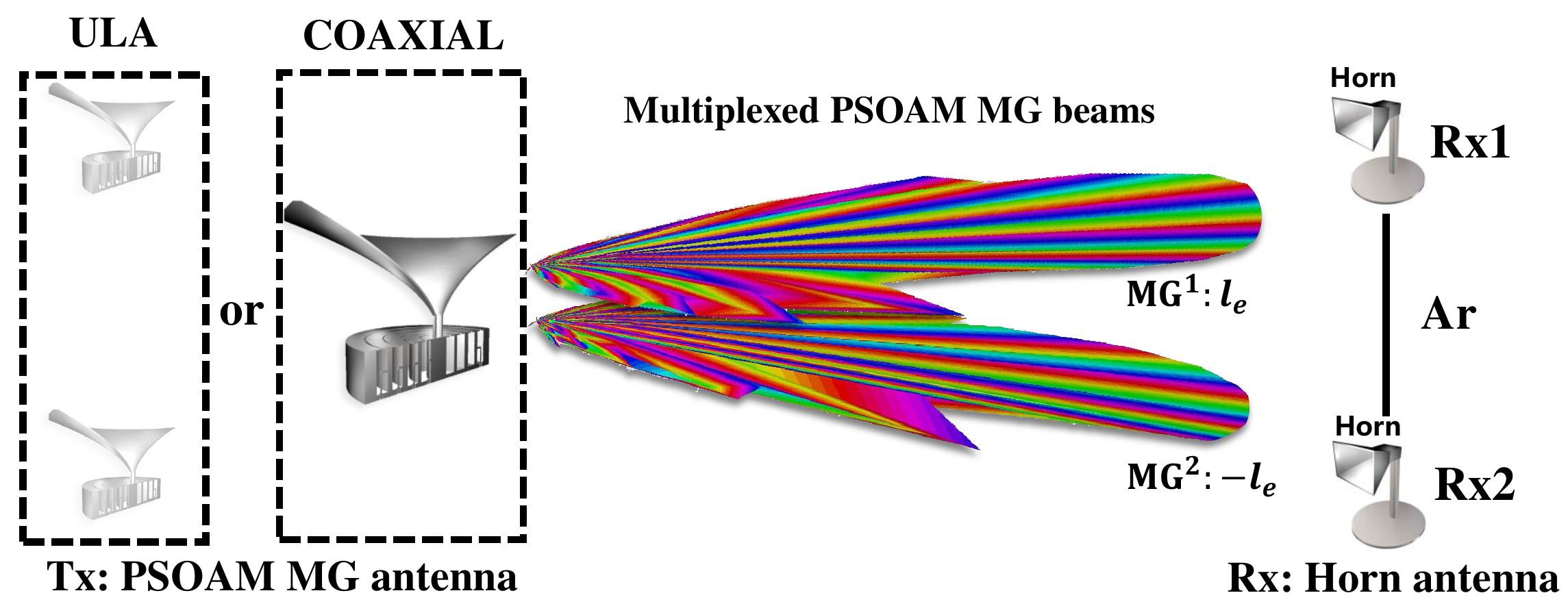}
	\caption{The sketch map of $2\times2$ PSOAM MG based LoS-MIMO communication link. MG's mainlobe possesses diversified and controllable phase information.}
	\label{fig1}	
\end{figure}

In spite of orthogonality between different OAM waves, it has been proved that OAM MDM system could be summed up as a subset of multiple-in-multiple-out (MIMO) technique\cite{edfors2011orbital}, but it still has a noteworthy superiority that signal processing complexity\cite{zhang2016mode} can be reduced. In addition, the phase singularity\cite{mohammadi2009orbital} and the inherent divergence\cite{padgett2015divergence} of conventional OAM wave make it harder to receive the beam within a whole angular aperture. At present, there is a newly proposed MIMO architecture\cite{zhang2016capacity,zhang2017experimental} that OAM antennas are merely applied to the transmitting end, which utilizes the diversity of OAM waves rather than the orthogonality. According to \cite{zhang2017experimental}, a special form of OAM wave called plane spiral OAM (PSOAM) wave has been adopted, each PSOAM wave propagates along the transverse direction ($\theta=90^\circ$) and the energy hole problem will no longer exist along the propagation axis. PSOAM wave's spiral phase distribution can provide a controllable degree of freedom (DoF) to decrease the spatial correlation between the sub-channels\cite{zhang2017experimental}. However, due to the omni-directivity in azimuthal direction of single mode PSOAM wave, most of the energy is wasted, which could lead to the deterioration of signal-to-noise (SNR) ratio at the receiving end. From the above, it's significant to find a feasible method to generate a directional beam, meanwhile, such a directional beam should still retain a spiral phase distribution within its mainlobe.

The beamforming can be realized by superposing eigenmodes of electromagnetic wave. Benefiting from the divergence consistency, PSOAM modes constitute a complete set of azimuthal eigenmodes. Superposing several specific PSOAM modes into a mode-group (MG), a reconfigurable beamforming scheme can be implemented \cite{zheng2017realization}. For conventional OAM waves, it's difficult to achieve that owing to the divergence inconsistency. PSOAM MG has its unique characteristics including directionality, vorticity and quasi-orthogonality\cite{zheng2017realization}. It has been demonstrated that superposing consecutive PSOAM modes into a MG can realize a high gain beam with a spiral phase distribution within its mainlobe. As for such a high gain beam, some researchers define it as \emph{quasi OAM wave}\cite{2019An,2018Quasi}. However, since it contains multiple OAM mode-components, so it would be more rigorous to call it \emph{mode-group}\cite{2020Direct}. PSOAM MG can be applied to the transmitting end to build the PSOAM MG based MIMO system (MG-MIMO)\cite{2020Performance}. Fig. \ref{fig1} shows the diagram of the proposed system, the transmitting end could adopt the coaxial arraying way or the uniform linear arraying (ULA) way. So far, MG has evoked other unique applications of spatial field direct-modulation\cite{shuang2020programmable,chen2020orbital} and low interception communications \cite{2019Low}. 

In this paper, the PSOAM MG based MIMO system has been experimentally verified for the first time, it's the proof-of-concept experiments under laboratory conditions. A compact multi-mode PSOAM antenna has been designed to generate multiple PSOAM MGs simultaneously\cite{zhu2020compact}, which can be used for building the array at the transmitting end. The bit-error-rate (BER) performance of the proposed system is measured and analyzed in a veritable LoS scenario. The experimental results prove that MGs can increase the SNR and provide another DoF to decrease the spatial correlation. Whereas, for conventional MIMO, the differences among sub-channels are merely caused by the phase difference that is brought about by different propagation paths. Moreover, we have also demonstrated a trial long-distance wireless 16-QAM transmission operated at 10.2 GHz, where the distance is 50 m (beyond the Rayleigh distance).
 
%This paper's structure arrangement is as follows: In Section II, .... In Section III,... In Section IV,.... Finally, this paper is concluded in Section V.

\section{PSOAM mode-group based MIMO communication}
\subsection{Channel model}
%\begin{table}[t]
%	\begin{center}
%		\caption{Parameters for the antenna}
%		\begin{tabular}{c|c}
%			\toprule[2pt]
%			\textbf{Symbols}&\textbf{Size (mm)}\\
%			\hline
%			a&78.60\\ 
%			b&2.80\\ 
%			c&0.68\\ 
%			d&WR-15\\ 
%			e&WR-15\\ 
%			f&0.70\\
%			\bottomrule[2pt]
%		\end{tabular}
%		\label{tab1}
%	\end{center}
%\end{table}
In the proposed MG-MIMO\cite{2020Performance}, the PSOAM MG antennas are merely applied to the transmitting end, the receiving end still use the common antenna like the standard horn antenna to form an ULA. The proposed architecture takes advantage of the phase diversity of the MG rather than the orthogonality just like most of the OAM MDM systems do. As shown in Fig. \ref{fig1}, the channel model is illustrated as a $2\times2$ LoS-MIMO system to make it clear. For more general condition, when it comes to $N$ MGs multiplexing, the number of horns should be equal to the number of multiplexing MGs. Based on the narrow-band MIMO theory\cite{goldsmith2005wireless,edfors2011orbital}, the general channel model of a $N\times N$ MG-MIMO system can be expressed as follows:
%{\setlength\abovedisplayskip{1pt}
%	\setlength\belowdisplayskip{1pt}
\begin{equation}
	\label{formula1}
	\begin{bmatrix}
	y_{1}\\
	y_{2}\\
	\vdots\\
	y_{N}
	\end{bmatrix}
	=
	\begin{bmatrix}
	h_{11}&h_{12}&\cdots&h_{1N}\\
	h_{21}&h_{22}&\cdots&h_{2N}\\
	\vdots&\vdots&\ddots&\vdots\\
	h_{N1}&h_{N2}&\cdots&h_{NN}
	\end{bmatrix}
	\begin{bmatrix}
	x_{1}\\
	x_{2}\\
	\vdots\\
	x_{N}
	\end{bmatrix}
	+
	\begin{bmatrix}
	n_{1}\\
	n_{2}\\
	\vdots\\
	n_{N}
	\end{bmatrix}
\end{equation}   
where $\mathbf{y}=[y_1,y_2,...,y_N]^T\in$ $\mathbb{C}^{N\times 1}$ is the complex receiving signal, $\left( \cdot\right) ^T$ is the matrix transpose operation; $\mathbf{x}=[x_1,x_2,...,x_N]^T\in$ $\mathbb{C}^{N\times 1}$ is the complex transmitting signal; $\mathbf{n}=[n_1,n_2,...,n_N]^T\in$ $\mathbb{C}^{N\times 1}$ is the complex additive white Gaussian noise vector\cite{lee2012digital} at the receiving end; and $\mathbf{H}=[h_{mn}]\in$ $\mathbb{C}^{N\times N}$ is the complex transfer matrix that consists of several transfer functions. Here, $\mathbf{H}$ can be either $\mathbf{H}^{\rm MIMO}$ or $\mathbf{H}^{\rm MG}$. To simplify the expression, we can write (\ref{formula1}) as $\mathbf{y}=\mathbf{H}\mathbf{x}+\mathbf{n}$ \cite{edfors2011orbital}.

The above-mentioned transfer matrix is made up of several transfer functions, given a propagation distance between a pair of transmitting and receiving antennas $d_{mn}$, the $\mathbf{H}^{\rm MIMO}$ can be expressed as  
\begin{equation}
	\label{formula2}
    \mathbf{H}^{\rm MIMO}
	=
	\beta_a
	\begin{bmatrix}
	\frac{\lambda e^{-jkd_{11}}}{4\pi d_{11}}&\cdots&\frac{\lambda e^{-jkd_{1N}}}{4\pi d_{1N}}\\
	\vdots&\ddots&\vdots\\
	\frac{\lambda e^{-jkd_{N1}}}{4\pi d_{N1}}&\cdots&\frac{\lambda e^{-jkd_{NN}}}{4\pi d_{NN}}
	\end{bmatrix}
\end{equation}   
where $\beta_{a}$ contains all the antenna performance parameter, $\lambda$ is the wavelength, $k=2\pi/\lambda$ represents the free space wave number, $\lambda /(4\pi d)$ refers to the path loss. All transfer functions in $\mathbf{H}^{\rm MIMO}$ mainly rely on the relative positions of transmitting and receiving array elements, which means that the differences between sub-channels are merely caused by different propagation paths. 

As for the MG-MIMO system, given a propagation distance $d_{mn}$ from the $n$th MG antenna to the $m$th horn, its transfer matrix $\mathbf{H}^{\rm MG}$ can be given as follows\cite{2020Performance}: 
\begin{equation}
\label{formula3}
	\mathbf{H}^{\rm MG}
	=
	\beta_a
	\begin{bmatrix}
	\frac{\lambda e^{-jkd_{11}}}{4\pi d_{11}}BP_{\rm{MG}^{1}}&\cdots&\frac{\lambda e^{-jkd_{1N}}}{4\pi d_{1N}}BP_{\rm{MG}^{N}}\\
	\vdots&\ddots&\vdots\\
	\frac{\lambda e^{-jkd_{N1}}}{4\pi d_{N1}}BP_{\rm{MG}^{1}}&\cdots&\frac{\lambda e^{-jkd_{NN}}}{4\pi d_{NN}}BP_{\rm{MG}^{N}}
	\end{bmatrix}
\end{equation}   
in (\ref{formula3}), $BP_{\rm{MG}^{n}}$ is the term of the $n$th MG's beam pattern. This term represents that the MG antenna is used for the transmitting end instead of the common antenna. $\rm MG^n$ $\left\lbrace l_f^n,l_f^n+\Delta l,...,l_f^n+(Q-1)\Delta l\right\rbrace$ is used to indicate a mode-group superposed by $Q$ PSOAM modes $l=l_f^n$, $l=l_f^n+\Delta l$,..., $l=l_f^n+(Q-1)\Delta l$. Its beam pattern in polar coordinate can be expressed below:
\begin{equation}
	\begin{aligned}
	BP_{\rm MG^{n}}\left ( \varphi  \right )&=\frac{1}{\sqrt{Q}}\sum_{q=0}^{Q-1}A_{q}e^{-j\left[ \left ( l_{f}^n+q\Delta l \right ) \varphi +\varphi _{0}^q \right]  }\\&\overset{\varphi _{0}^q=0^{\circ}, A_{q}=1}{=}\frac{1}{\sqrt{Q}}{\textcolor[rgb]{0,0.25,0.53}{\frac{\sin(\frac{Q\varphi}{2}\Delta l)}{\sin(\frac{\varphi}{2}\Delta l)}}}{\textcolor[rgb]{1,0,0}{e^{-j(l_{f}^n+\Delta l \frac{Q-1}{2})\varphi}}}
	\end{aligned}
	\label{formula4}	
\end{equation}
where $A_{q}$ and $\varphi _{0}^q $ is the magnitude and initial phase of the $q$th PSOAM mode, respectively; $l_f^n$ denotes the first PSOAM mode among the $\rm MG^n$; $\Delta l$ refers to the mode interval; and $\varphi$ is the azimuthal angle; the constant $1/\sqrt{Q}$ is aimed at normalizing the power for a fairer comparison with conventional MIMO. For the situation of equi-amplitude and in-phase, the beam pattern expression in (\ref{formula4}) can be extended to the product of the {\color[rgb]{0,0.25,0.53}{amplitude term}} and the {\color[rgb]{1,0,0}{phase term}}. The {\color[rgb]{0,0.25,0.53}{amplitude term}} characterizes the directionality of a MG, which is related to $Q$ and $\Delta l$. To realize a high gain pencil beam, the mode interval $\Delta l$ should be equal to $1$, i.e. PSOAM modes in a MG must be consecutive modes. The {\color[rgb]{1,0,0}{phase term}} describes the spiral phase distribution within MG's mainlobe, which is not only about $Q$ and $\Delta l$ but also about $l_{f}^n$. It's worth noticing that PSOAM MG can achieve azimuthal radiation pattern diversity\cite{zheng2017realization}: 
\begin{enumerate}[i)]
	\item{\textbf{Directionality}}: The PSOAM modes and the azimuthal angle constitute a \emph{Fourier} transformation pair\cite{jack2008angular}, a MG with a high gain beam has more components in the OAM mode spectrum, i.e. the beamwidth of a MG decreases with increasing $Q$\cite{zheng2017realization}.
	\item {\textbf{Vorticity}}: The phase distribution within MG's mainlobe is still linear to the azimuthal angle. The phase slope within the mainlobe represents the vorticity of a MG. Here, we may define $l_{e}^n$ as the equivalent OAM order of a $\rm MG^n$, which can be calculated from the phase term in (\ref{formula4}):
	\begin{equation}
		\label{formula5}
		l_e^n=l_{f}^n+\Delta l \frac{Q-1}{2}
	\end{equation}
\end{enumerate}

When both the transmitting and receiving end obtain the channel state information (CSI)\cite{yu2020analyze}, i.e. once getting the transfer matrix shown in (\ref{formula2}) or (\ref{formula3}), the information-theoretic channel capacity limit can be calculated utilizing singular value decomposition (SVD) algorithm\cite{2020Performance}:
%{\setlength\abovedisplayskip{6pt}
%	\setlength\belowdisplayskip{2pt}
\begin{equation}
	\begin{split}
		\label{formula6}
		C=\sum_{k=1}^{rank(\mathbf{H})}\log_{2}\left(1+\frac{\nu_{k}^2 P_{k}}{\sigma_{n}^2} \right)\ (bits/s/Hz)\\
		s.t.\quad P_{total}=\sum_{k=1}^{rank(\mathbf{H})}P_k
	\end{split}
\end{equation}
where $\nu_{k}\in\left\lbrace \nu_{1},\nu_{2},...,\nu_{rank(\mathbf{H})}\right\rbrace$ refers to the singular value decomposed by $\mathbf{H}$; $\sigma_{n}^2$ is the noise variance; and $rank\left( \cdot\right)$ represents the rank of a matrix. $P_{total}$ is the total transmitting power, $P_k$ is the power transmitted in the sub-channels, where an optimized power allocation scheme called \emph{water-filling principle}\cite{yu2004iterative} can be adopted when CSI has been obtained. Given a receiving SNR, $P_{total}$ can be estimated using the transfer function. After that, the capacity gain (CG) of MG-MIMO relative to single-in-single-out (SISO) system\cite{edfors2011orbital} could be simulated numerically.
\begin{table}[t]
%	\vspace*{-1.0mm}
	\begin{center}
		\caption{Simulation Parameters}
		\begin{tabular}{c|c c c}
			\toprule[2pt]
			\textbf{Channel models}&\textbf{MIMO}&\textbf{PSOAM-MIMO}&\textbf{MG-MIMO}\\
			\hline
			$l_{max}$&\multicolumn{3}{|c}{20} \\ 
			$f_c$$^{\rm a}$&\multicolumn{3}{|c}{10 GHz} \\ 
			Receiving SNR&\multicolumn{3}{|c}{30 dB} \\ 
			\hline
			$A_{t}$$^{\rm b}$&$l_{max}\lambda/\pi$&Coaxial&Coaxial\\
			$A_{r}$$^{\rm c}$&$l_{max}\lambda/\pi$&$l_{max}\lambda/\pi$& $l_{max}\lambda/\pi$\\
			\bottomrule[2pt]
		\end{tabular}\\
		\vspace*{+1.0mm}
		\footnotesize{$^a$$f_c$ is the carrier frequency; $^b$$A_t$ is the transmitting aperture; and $^c$$A_t$ is the receiving aperture.}\\
		\label{tab1}
	\end{center}
\end{table}
\begin{figure}[t]
%    \vspace*{-6mm}
	\centering
	\includegraphics[width=2.6in]{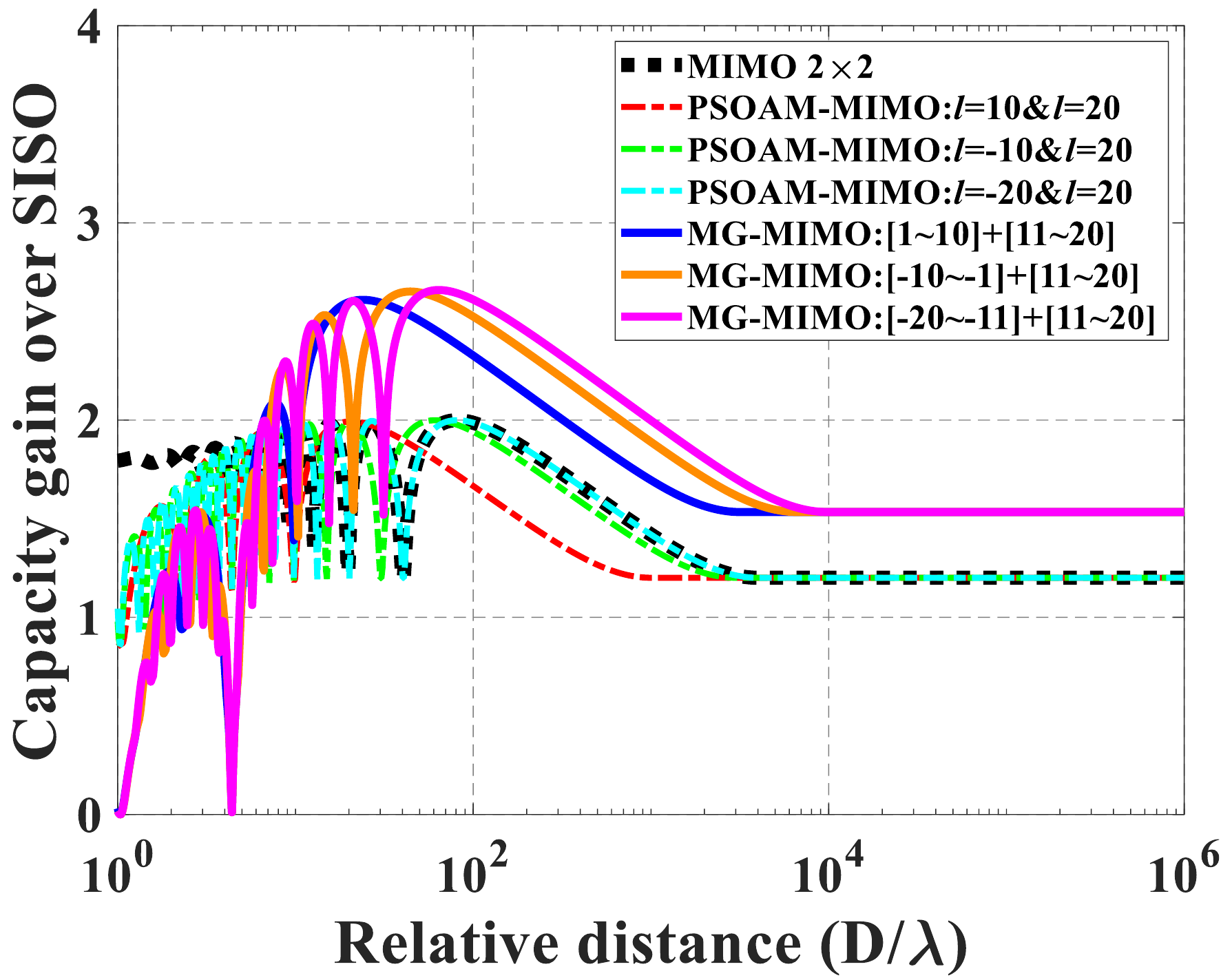}
	\caption{CG over SISO system for conventional MIMO, PSOAM-MIMO and MG-MIMO at the size $2\times2$. The receiving SNR is 30 dB. Rayleigh distance\cite{balanis2016antenna} of $2A_{r}^2/\lambda$ is approximately $81.1\lambda$.}
	\label{fig2}
\end{figure}

Fig. \ref{fig2} presents the CG of conventional MIMO \cite{edfors2011orbital}, PSOAM based MIMO (PSOAM-MIMO)\cite{zhang2017experimental} and MG-MIMO in a LoS scenario, where the transmitting end of all the latter two systems take the coaxial arraying way. Pivotal simulation parameters are given in Table \ref{tab1}, where $l_{max}$ is the maximum  order PSOAM mode among multiplexing MGs; and $l_{max}\lambda/\pi$ is the aperture of MG antenna, which makes sure to generate the $l$th order PSOAM mode in theory\cite{2020Performance}. The multiplexed MGs are $\rm MG^{1}$ $\left \{1,2,...,9,10\right \}$, $\rm MG^{2}$ $\left \{11,12,...,19,20\right \}$, $\rm MG^{3}$ $\left \{-10,-9,...,-2,-1\right \}$ and $\rm MG^{4}$ $\left \{-20,-19,...,-12,-11\right \}$. Each MG has the same certain directionality\cite{zheng2017realization}, however, $l_e^n$ of each MG is 5.5, 15.5, -5.5 and -15.5, respectively. In addition, the CG is simulated numerically at a receiving SNR of 30 dB no matter how long the communication distanc $D$ is. 

The CG reaches its peak ever faster after a limited $D$, and then it will deteriorate dramatically beyond the rayleigh distance ($2A_{r}^2/\lambda$). Benefitting from the SNR improvement that MG's directionality brings, MG-MIMO has a better peak CG about 2.6. Due to the consistent beam directionality of four MGs\cite{zheng2017realization}, the peak CG for the case of three pairs of multiplexed MGs remain the same. To obtain a specific capacity gain such as 2, MG-MIMO can achieve a longer propagation distance, which is raised from $84\lambda$ of MIMO to $328\lambda$ (blue solid line,  $\Delta l_e^{12}=10$), $655\lambda$ (orange solid line, $\Delta l_e^{23}=21$) and $993\lambda$ (magenta solid line,  $\Delta l_e^{24}=31$), respectively. Notably, for the same $A_r$ and same directionality, if the difference value of equivalent OAM order $\Delta l_e$ between multiplexed MGs is larger, the correlation of sub-channels will be lower, which verifies numerically that MGs provide a controllable DoF to decrease the spatial correlation. 

As for the PSOAM-MIMO, without the SNR improvement, the peak CG of PSOAM-MIMO is the same as MIMO system about 2. As the difference value of OAM order $\Delta l$ between multiplexed PSOAM waves increases, the propagation distance will also increase. However, because the sub-channel difference caused by coaxially arraying PSOAM waves is weaker than that caused by the propagation path, its communication distance is even shorter than MIMO for a certain CG. If the transmitting end of PSOAM-MIMO adopts the ULA arraying way, it would show a better performance of communication distance\cite{zhang2017experimental}.

\subsection{The transmitting end: PSOAM MG antenna}

\begin{figure}[t]
	%	\vspace*{-4mm}
	\centering
	\includegraphics[width=3.4in]{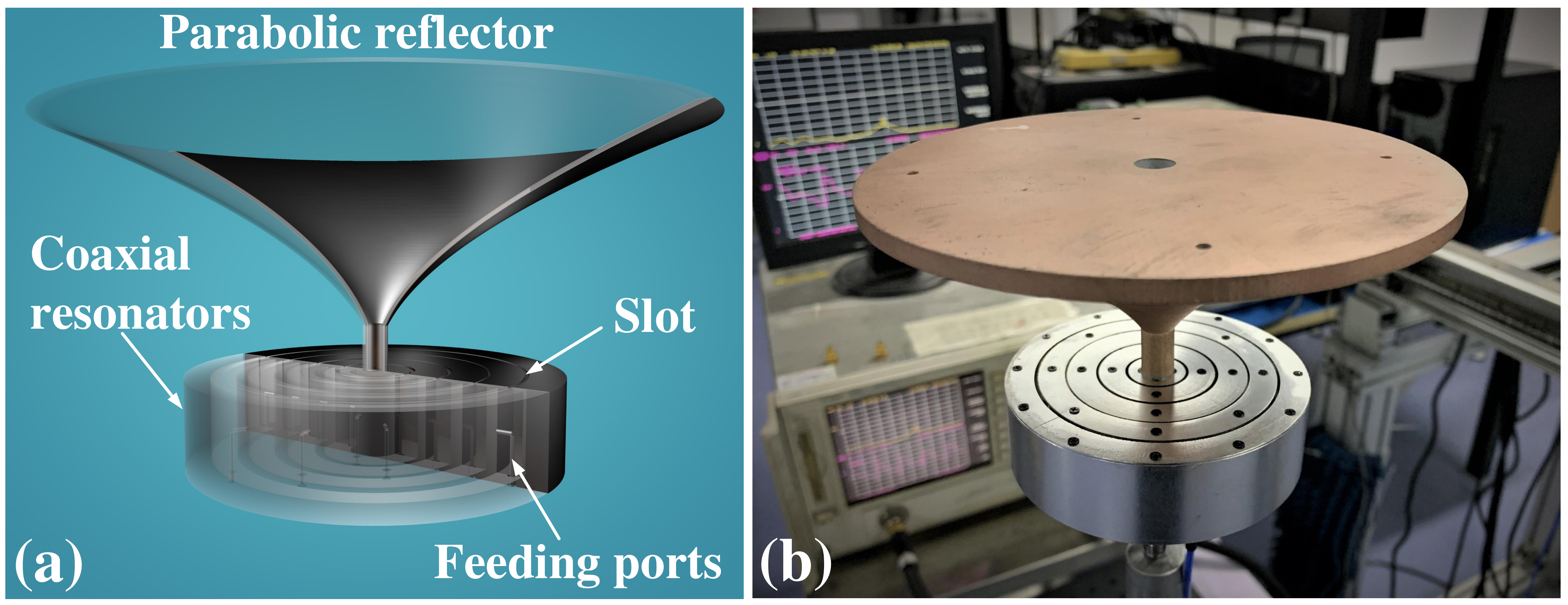}
	\caption{(a) The cross-view of the mentioned PSOAM MG antenna. (b) A picture of the fabricated antenna.}
	\label{fig3}
\end{figure}

For the sake of constructing the proposed MG-MIMO communication link, a PSOAM MG antenna that can simultaneously radiate several required MGs needs to be manufactured. To achieve such a goal, a compact multi-mode PSOAM antenna has been demonstrated\cite{zhu2020compact}. The proposed antenna is made up of four coaxial resonators and a parabolic reflector. The cross-sectional view and the real picture of the antenna are shown in Fig. \ref{fig3}. Its physical aperture (the diagonal length of cross-section) is approximately 170 mm (about $5.8\lambda$). 

\begin{figure}[t]
	%	\vspace*{-4mm}
	\centering
	\includegraphics[width=3.5in]{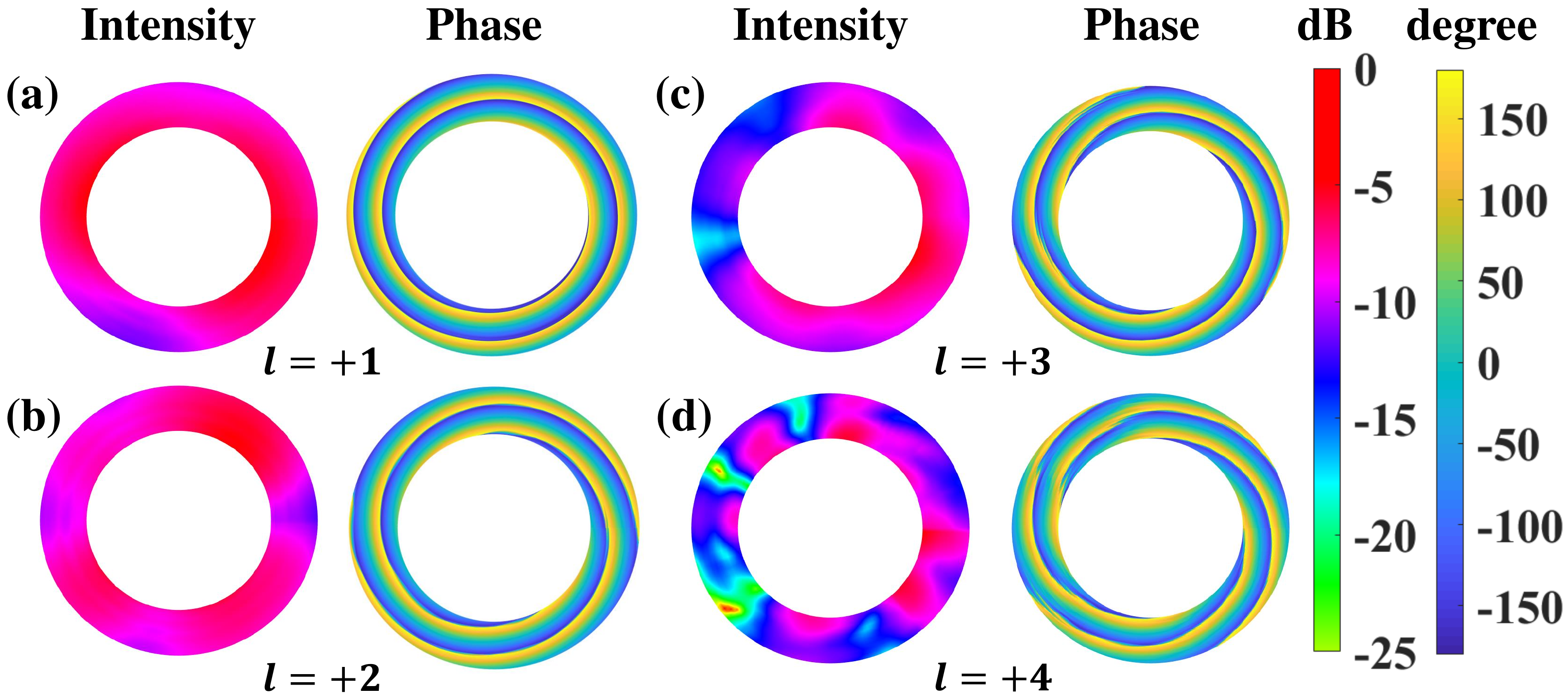}
	\caption{The measured near-field results including the phase and the normalized intensity distributions in the transverse plane of PSOAM waves (a) $l=1$, (b) $l=2$, (c) $l=3$ and (d) $l=4$.}
	\label{fig4}
\end{figure}

In virtue of the $90^\circ$ hybrid coupler\cite{pozar2011microwave}, in the same resonator, the $\pm l$ order PSOAM waves can be excitated concurrently\cite{hui2015multiplexed}. Therefore, such an antenna could synchronously generate 8 PSOAM waves with the OAM order $l=\pm 1, \pm 2, \pm 3$ and $\pm 4$, and in the following experiments, all the PSOAM modes are used at 10.2 GHz. Only the four typical examples are presented, their measured phase and normalized intensity distributions in the transverse plane are illustrated in Fig. \ref{fig4}. The results display the distinct feature of single mode PSOAM wave. Obviously, the central singularity of PSOAM waves will exist inside the antenna itself, the energy hole problem can be reasonably avoided. The phase changes $2\pi l$ along the circumference. The inhomogeneity of the intensity is caused by the errors of manufacture and the energy loss of resonant cavity\cite{pozar2011microwave,li2015radiation}, the latter factor results in the more serious inhomogeneity for the higher order PSOAM mode.

\begin{figure}[t]
	%	\vspace*{-4mm}
	\centering
	\includegraphics[width=3.5in]{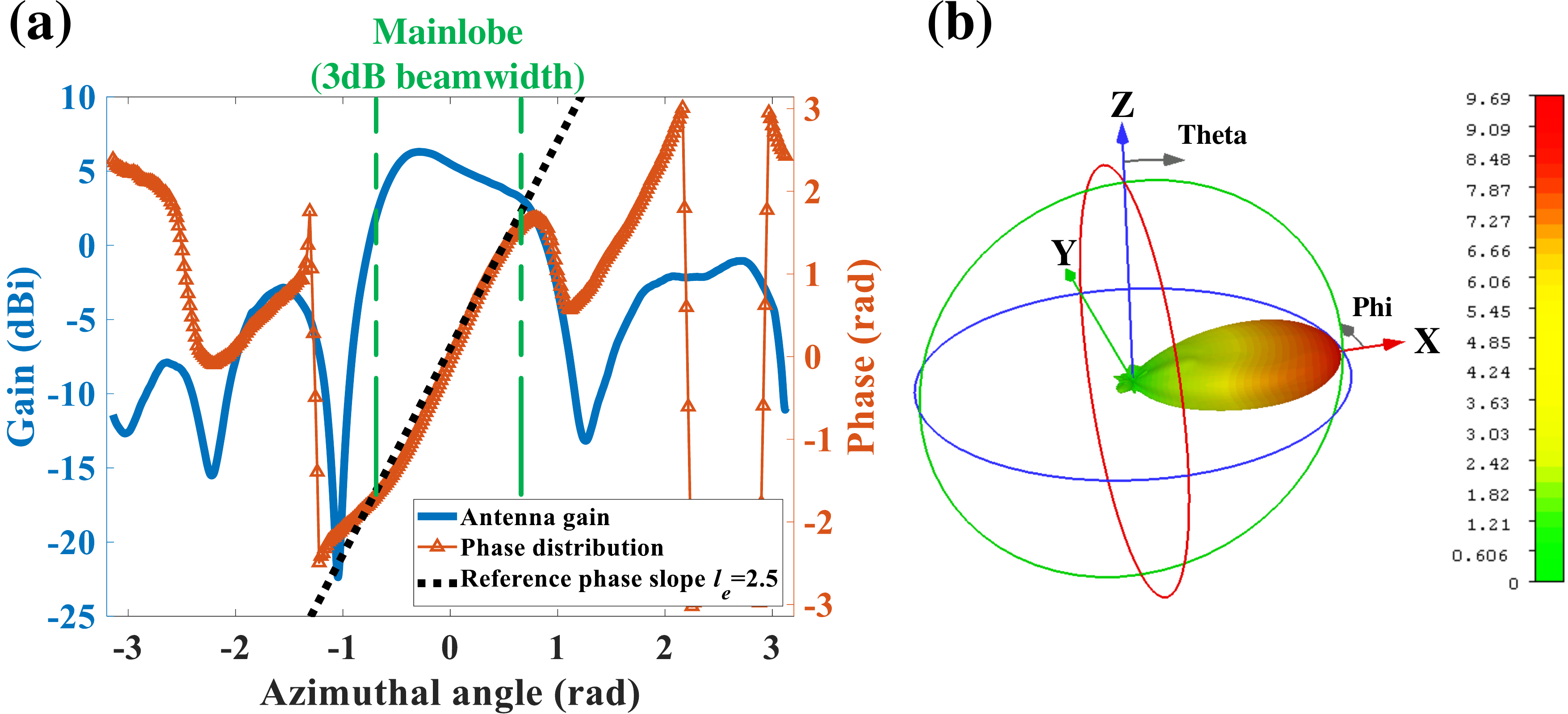}
	\caption{(a) The measured far-field directivity (blue) and phase distribution (dark-orange) for $\rm MG$ $\left \{1,2,3,4\right \}$ at $XOY$ plane ($\theta=90^\circ$). (b) The simulated 3D far-field pattern (linear scale) for $\rm MG$ $\left \{1,2,3,4\right \}$.}
	\label{fig5}
\end{figure}

Besides, each PSOAM wave can be manipulated independently relying on an amplitude-phase controllable feeding network, which can realize the generation of MG. Fig. \ref{fig5}(a) shows the measured far-field directivity and phase distribution of $\rm MG$ $\left \{1,2,3,4\right \}$ at $\theta=90^\circ$ plane. The mainlobe magnitude and 3dB beamwidth of azimuthal angle are $6.3$ dBi and $71^\circ$, respectively. The measured phase within the mainlobe presents a linear distribution with respect to azimuthal angle. Therefore, it proves that $\rm MG$ possesses the directionality and the vorticity at the same time. The linearity of phase slope is about good, the estimation value of $l_e$ is 2.5, which is in good agreement with the theoretical reference value (black). In addition, the simulated 3D far-field pattern is shown in Fig. \ref{fig5}(b), it can be seen more visually that a high gain pencil beam can be generated in the transverse plane. Additional design parameters and electrical performance about this antenna can be found in \cite{zhu2020compact}.

\begin{figure*}[t]
	%	\vspace*{-4mm}
	\centering
	\includegraphics[width=6.3in]{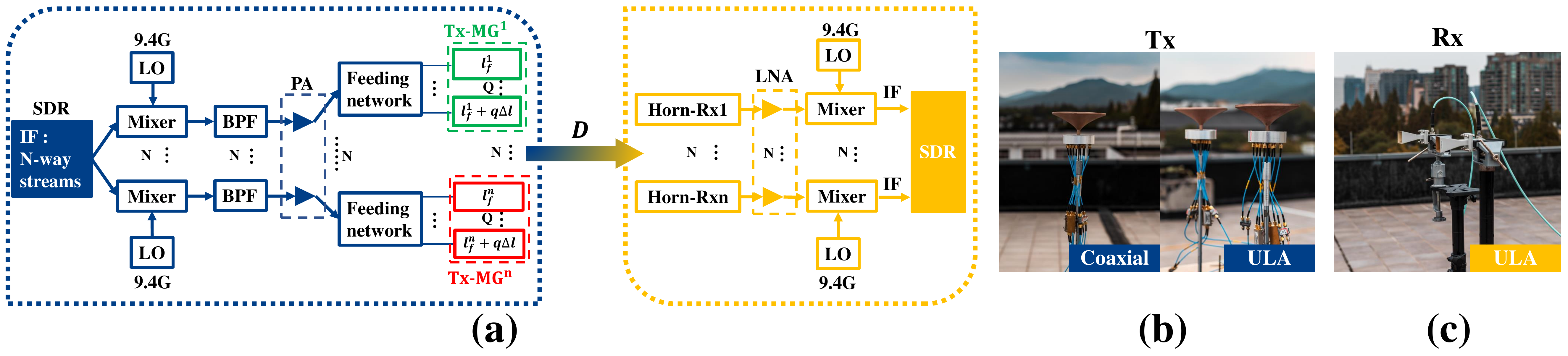}
	\caption{(a) The schematic diagram of a $N\times N$ MG-MIMO communication architecture in a LoS scenario with a communication distance $D$. The $N$ PSOAM MGs carrying $N$ data streams are used for the transmitting end, in which the carrier frequency $f_c$ is 10.2 GHz. SDR: software defined radio. IF: intermediate frequency signal. LO: local oscillator. BPF: band pass filter. PA: power amplifier. Tx/Rx: transmitting antenna and receiving antenna. Horn: standard gain horn antenna. LNA: low noise amplifier. Taking a $2\times 2$ MG-MIMO communication link as an example, (b) the transmitting end can adopt the coaxial arraying way or the ULA arraying way utilizing the proposed MG antennas; (c) the receiving end adopts the ULA arraying way utilizing the horn antennas.}
	\label{fig6}
\end{figure*}
\subsection{Communication system architecture}
A schematic overview of a $N\times N$ PSOAM MG based MIMO system is shown in Fig. \ref{fig6}(a). For the transmitting module, in virtue of orthogonal frequency division multiplexing (OFDM)\cite{goldsmith2005wireless} technique with 56 sub-carriers, $N$-way independent modulated baseband signals with the 20 MHz bandwidth (BW) whose intermediate frequency (IF) is set as 800 MHz are generated by a software defined radio (SDR, V3 Technology YunSDR Y410s)\cite{mitola2002cognitive} platform. The modulation formats are quadrature phase shift keying (QPSK), 16-Quadrature Amplitude Modulation (QAM) and 64-QAM\cite{lee2012digital}. $N$-way IF signals are fed to the IF ports of $N$ mixers, $N$-way 9.4 GHz continuous wave (CW) signals (local oscillator, LO) are fed to the LO ports of $N$ mixers, so far, the baseband signals are up-converted to the carrier frequency $f_c$ of 10.2 GHz and these radio frequency signals are exported from the RF ports of $N$ mixers. Then, after getting through the band pass filters (BPF) and power amplifiers (PA), $N$-way RF signals carrying data streams are fed to the amplitude-phase controllable feeding networks. The feeding network consists of a power divider, adjustable phase shifters and adjustable attenuators\cite{zhu2020compact}, which is used to manipulate the intensity and the initial phase of each PSOAM wave among a MG. Combining the feeding networks and the proposed MG antennas, $N$ MGs carrying $N$-way independent data streams 
can be radiated and multiplexed in free space. Owing to that the MG antenna can simultaneously generate 8 different PSOAM modes, these modes could be divided into multiple MGs, hence it's possible for the transmitting end to adopt the coaxial arraying way in addition to the ULA arraying way, Fig. \ref{fig6}(b) shows the experimental setup. In a sense, the coaxial arraying way can be considered that $N$ multiplexed MGs radiate from the origin in a polar coordinate.  

After a free space transmission in a LoS scenario, $N$-way RF signals carrying data streams reach the receiving horn antennas. On account of the pathloss, in order to meet the certain SNR requirement, the receiving signals need to be amplified by the low noise amplifiers (LNA). Following by passing through the mixers, the RF signals are down-converted to the IF signals (800 MHz). Then, the IF signals are fed to the SDR platform to realize the process of signal demultiplexing, in which the second amplification and the second down-conversion are operated to get the digital baseband signals. Finally, the transmitted modulated signals can be recovered, the bits-error rate (BER), received constellation diagrams and error vector magnitude (EVM) performance could be analyzed. Importantly, the MIMO demultiplexing algorithm used in  SDR platform is \emph{Zero-Forcing}\cite{spencer2004zero} algorithm. Besides, as shown in Fig. \ref{fig6}(c), the receiving end takes the ULA arraying way to form an antenna array utilizing standard gain horn antennas.

\begin{figure}[t]
	%	\vspace*{-4mm}
	\centering
	\includegraphics[width=2.8in]{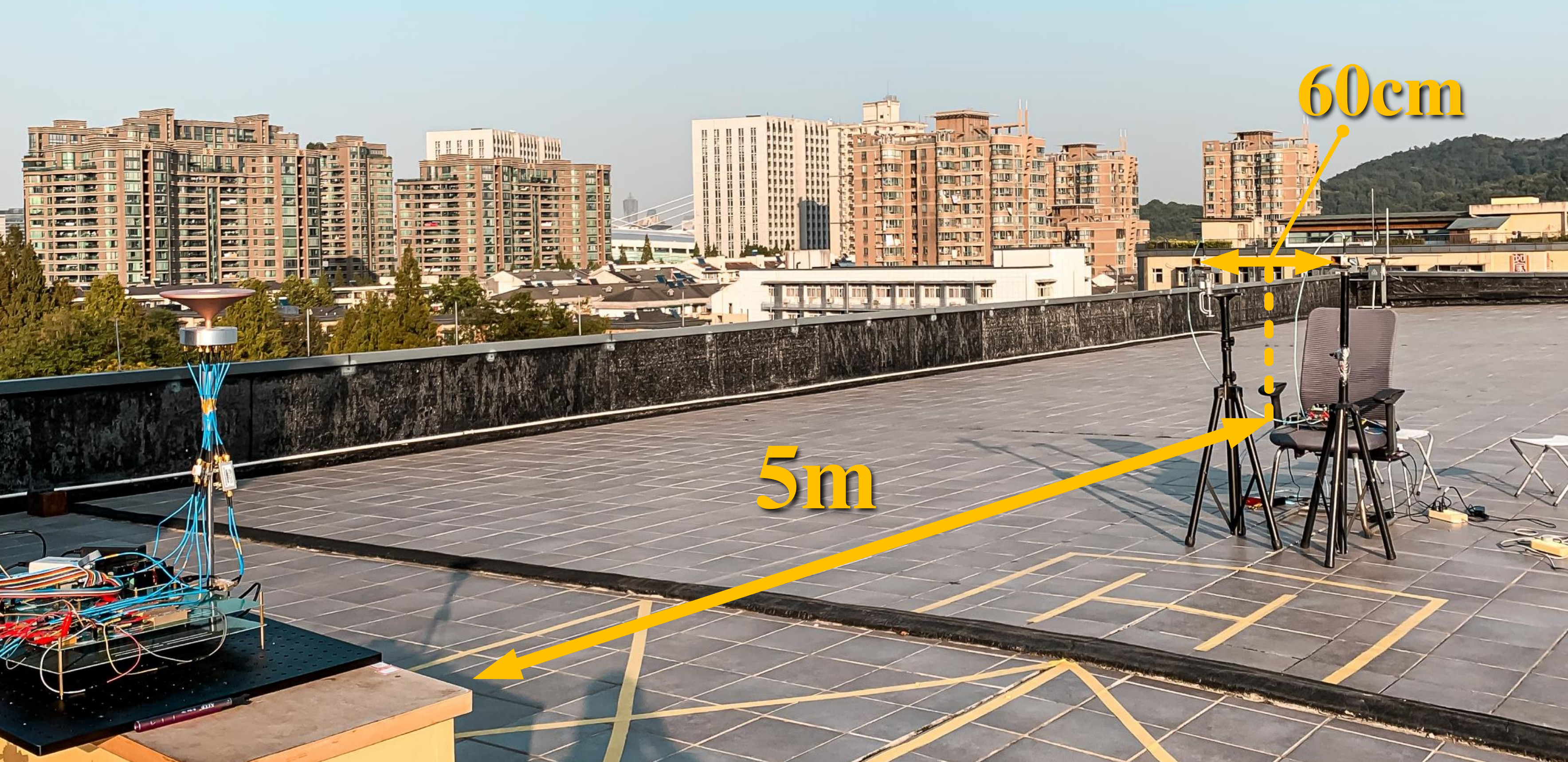}
	\caption{The experimental setup used to study the impact on communication link performance of $\rm MG$'s vorticity.}
	\label{fig7}
\end{figure}
\section{proof-of-principle experiments}
Limited by the number of channels available for our SDR platform, in  following several experiments, we set up a $2\times 2$ MG-MIMO  prototype to demonstrate and analyze the related communication performance.
\subsection{Impact on link performance of MG's vorticity}
It has been verified the validity numerically that, for the same $A_r$ and same directionality of MGs, if $\Delta l_e$ between multiplexed MGs is larger, the correlation of sub-channels will be lower. To eliminate the sub-channel difference caused by the different propagation paths, the transmitting end adopts the coaxial arraying way, here, the sub-channel difference is merely brought by the linear phase distribution of MGs. At the receiving end, two horn antennas are placed at the symmetry positions along the
propagation axis. The aperture of the receiving ULA $A_r$ is set to 0.6 m, and the communication distance $D$ is 5 m, as illustrated in Fig. \ref{fig7}. In addition, the total number $Q$ of PSOAM modes in each $\rm MG$ needs to stay the same\cite{zheng2017realization} and therefore the beam gain (directionality) of each MG can stay almost the same, which makes sure that the receiving SNR can remain consistent for each control group. Four pairs of MGs used for conducting the comparison experiments are $\rm MG^1 \left\lbrace -1,1\right\rbrace$ and $\rm MG^2 \left\lbrace -2,2\right\rbrace$ with $\Delta l_e=0$ (a pair of plane waves\cite{zheng2017realization}), $\rm MG^3 \left\lbrace 1,2\right\rbrace$ and $\rm MG^4 \left\lbrace -2,-1\right\rbrace$ with $\Delta l_e=3$, $\rm MG^5 \left\lbrace 2,3\right\rbrace$ and $\rm MG^6 \left\lbrace -3,-2\right\rbrace$ with $\Delta l_e=5$, $\rm MG^7 \left\lbrace 3,4\right\rbrace$ and $\rm MG^8 \left\lbrace -4,-3\right\rbrace$ with $\Delta l_e=7$, respectively. 
\begin{figure}[t]
%    \vspace*{-4mm}
	\centering
	\includegraphics[width=3.4in]{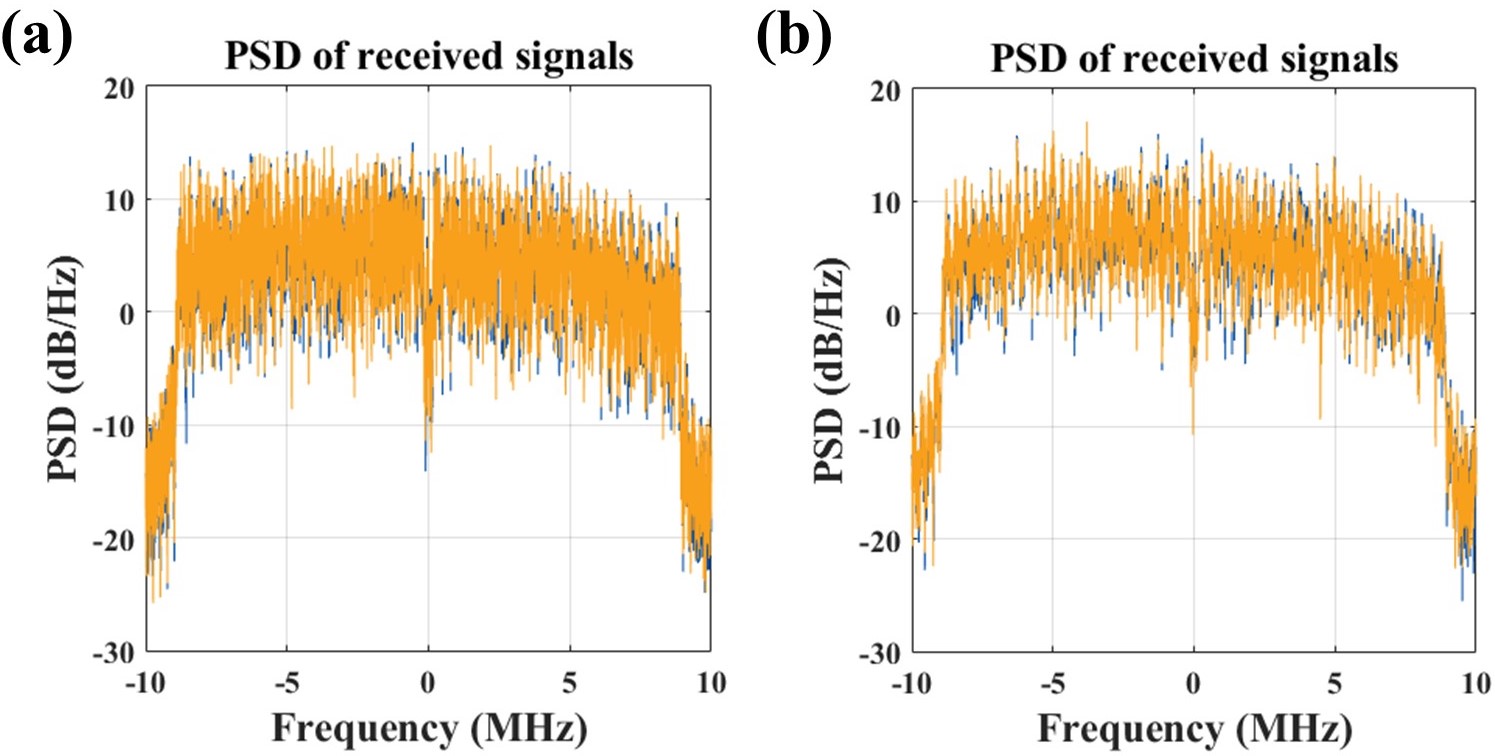}
	\caption{The typical PSD of the received signals (a) at a receiving SNR of 28 dB in the QPSK transmission link and (b) at a receiving SNR of 30 dB in the 16-QAM transmission link.}
	\label{fig8}
\end{figure}
\begin{figure*}[t]
	%	\vspace*{-4mm}
	\centering
	\includegraphics[width=5.6in]{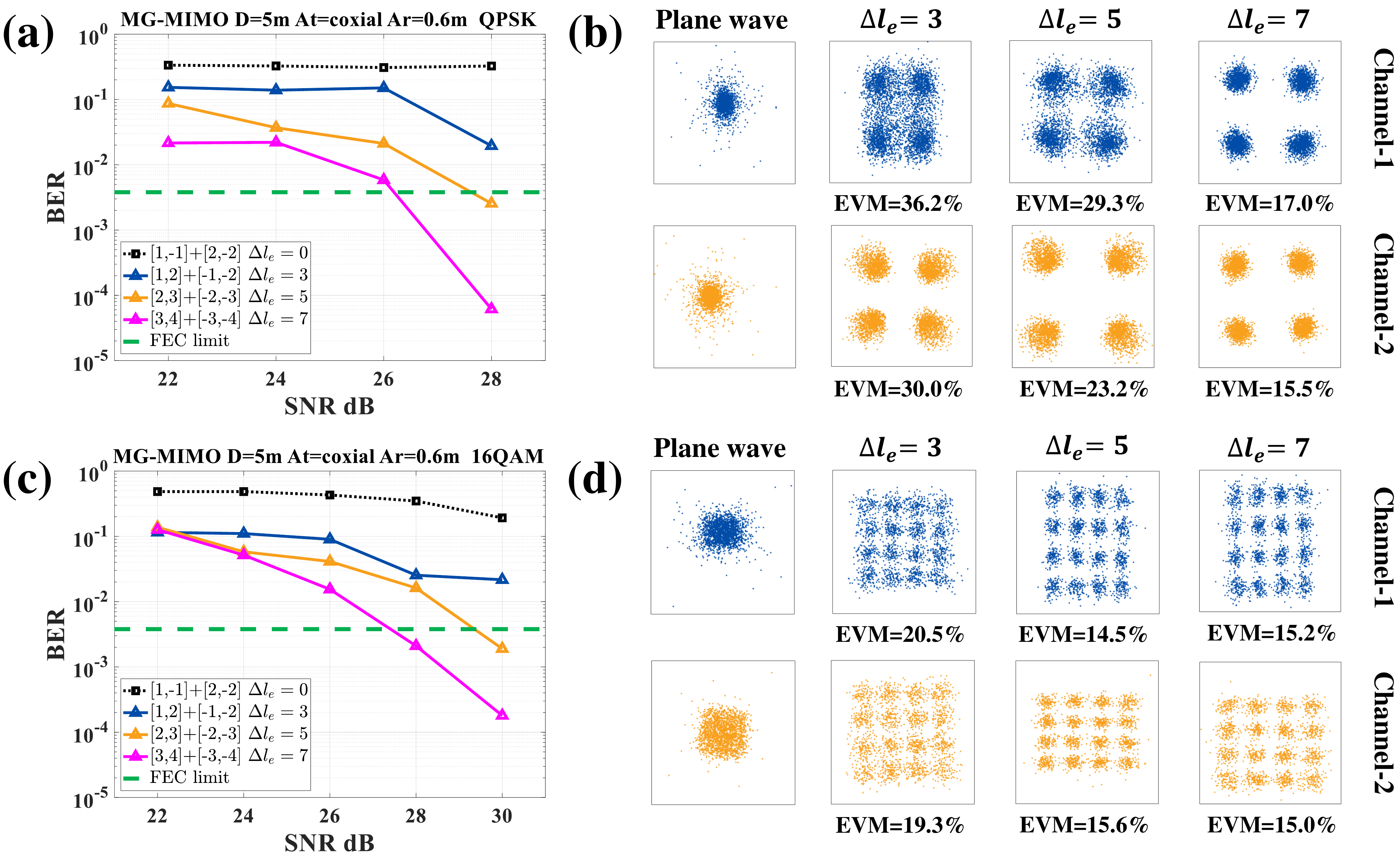}
	\caption{Experimental results for analyzing the impact on communication link performance of MG's vorticity. The QPSK transmission: (a) the measured raw BER performance and (b) the typical received constellations with the EVM at a receiving SNR of 28 dB for these four pairs of multiplexed MGs. Likewise, the 16-QAM transmission: (c) the measured BER and (d) the received constellations with the EVM at a receiving SNR of 30 dB.}
	\label{fig9}
\end{figure*}

To observe the characteristic in frequency domain of the received signals, Fig. \ref{fig8}(a) and Fig. \ref{fig8}(b) show the typical received power spectral density (PSD) of the QPSK transmission at a SNR of 28 dB and the 16-QAM transmission at a SNR of 30 dB, respectively. Such the spectrum can be achieved by using the fast \emph{Fourier} transformation (FFT) to do the calculation of the received baseband signal, where the signal is sampled by the 12-bit analog-to-digital (ADC) in our SDR platform. As we can see, the central sub-carrier of a typical OFDM symbol with the BW of 20 MHz  
is set to the null in order to overcome the problem of direct-current (DC) shift\cite{5307322,zhao2019programmable}. Here, an OFDM symbol contains 56 sub-carriers, of which 52 sub-carriers can be used for the data transmission. The sub-carrier interval is 312.5 KHz. 

The measured raw BER curves of the QPSK transmission and the 16-QAM transmission have been shown in Fig. \ref{fig9}(a) and Fig. \ref{fig9}(c), respectively. Fig. \ref{fig9}(b) shows the typical received constellations with the EVM performance of the QPSK transmission at a receiving SNR of 28 dB for each pair of multiplexed MGs. Similarly, Fig. \ref{fig9}(d) shows the corresponding results in the case of 16-QAM transmission at a receiving SNR of 30 dB. When the receiving aperture $A_r$, the total number $Q$ and the communication distance $D$ are respectively fixed as 0.6 m, 2 and 5 m, for both two modulation formats, as $\Delta l_e$ increases from 0, 3, 5 and 7, the corresponding BER value is reduced to a certain extent, i.e. the communication performance is getting better for the case of larger $\Delta l_e$. It's important and worthy of attention that, here, the difference of sub-channels is entirely caused by the MG's linear phase distribution rather than the propagation path difference like MIMO technique does. Due to the coaxial arraying way at the transmitting end, for the receiving horns arranged symmetrically along the propagation axis, the wave-path of two plane waves carrying data streams to the two receiving antennas are almost the same, which   
means that there is no spatial diversity\cite{cho2010mimo} between the multiplexed signals. At this time, the system cannot support the multiplexing, severe aliasing occurs between the multiplexed signals, its received constellations are shown in Fig. \ref{fig9}(b) and Fig. \ref{fig9}(d). As a consequence, for the case of $\Delta l_e=0$, its BER values at each SNR for both modulation formats are all above the level of $20\%$, it's an intolerable level in wireless communication system\cite{goldsmith2005wireless}. 

In the case of the QPSK transmission, as far as the test results are concerned, when $\Delta l_e$ is 5 or 7, the MG-MIMO system can reach a raw BER level of $3.8\times10^{-3}$, which is a level allowing extremely low error rates in virtue of efficient forward error correction (FEC) codes\cite{lee2012digital}. Benefiting from the lower sub-channel correlation, the multiplexing of $\rm MG^7$ and $\rm MG^8$ can reach the FEC limit at a lower SNR of about 26 dB, and the multiplexing of $\rm MG^5$ and $\rm MG^6$ needs a 28 dB SNR to reach the same limit. At the same receiving SNR of 28 dB, the EVM performance of two channels for $\Delta l_e=3$, $\Delta l_e=5$ and $\Delta l_e=7$ are $36.2\%$ and $30.0\%$, $29.3\%$ and $23.2\%$, $17.0\%$ and $15.5\%$, respectively.

As for the 16-QAM transmission, similarly, when $\Delta l_e$ is 5 or 7, the MG-MIMO system can reach the FEC limit. The multiplexing case of $\Delta l_e=7$ reaches the FEC limit at a lower SNR of about 28 dB, and the multiplexing case of $\Delta l_e=5$ requires a higher SNR of 30dB to reach this limit. At a certain receiving SNR of 30 dB, the EVM performance of two channels for $\Delta l_e=3$, $\Delta l_e=5$ and $\Delta l_e=7$ are $20.5\%$ and $19.3\%$, $14.5\%$ and $15.6\%$, $15.2\%$ and $15.0\%$, respectively.

\subsection{Impact on link performance of MG's directionality}
\begin{figure}[t]
	%    \vspace*{-4mm}
	\centering
	\includegraphics[width=2.5in]{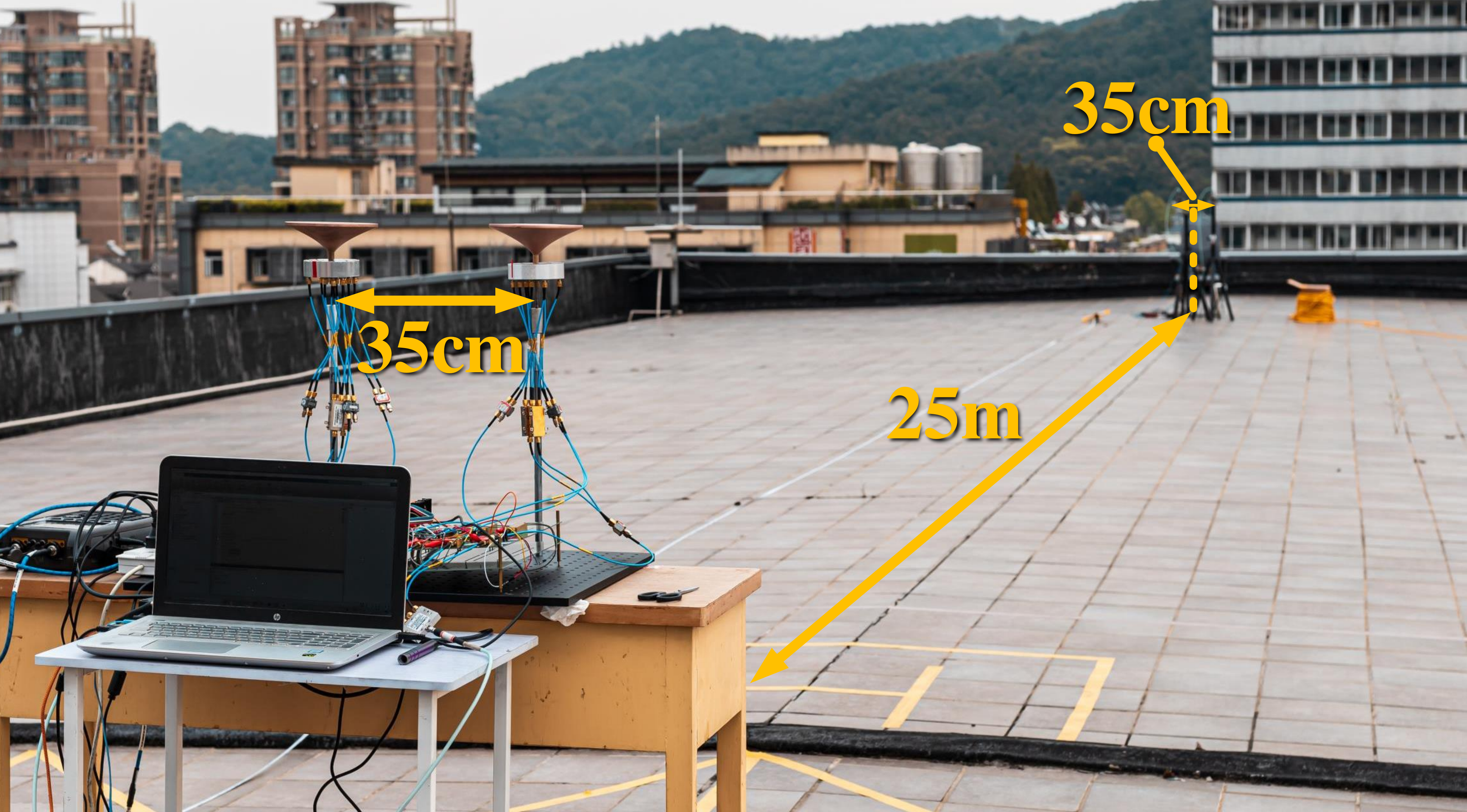}
	\caption{The experimental setup used to study the impact on communication link performance of $\rm MG$'s directionality.}
	\label{fig10}
\end{figure}

The PSOAM-MIMO system\cite{zhang2017experimental} also utilizes the diversity of PSOAM waves rather than the orthogonality, where the diversity is given by PSOAM wave's azimuthal spiral phase distribution. Unfortunately, because of the omni-directivity (as shown in Fig. \ref{fig4}) in azimuthal direction of single mode PSOAM wave, most of the energy is wasted, which will result in the deterioration of the receiving SNR. The beam generated by the PSOAM MG technique can realize the azimuthal beamforming to achieve the convergence of energy. In order to compare MG-MIMO with PSOAM-MIMO, to refer to the experimental setup in \cite{zhang2017experimental}, the transmitting and receiving end will employ the ULA arraying way, $A_t$ and $A_r$ are all set as 0.35 m. In order to fully demonstrate the beam gain superiority of MG over single mode PSOAM wave, as shown in Fig. \ref{fig10}, the communication distance $D$ of this outdoor experiment is set to be longer, where $D$ is 25 m. To strictly control the variable, the mode interval $\Delta l$ between multiplexed single mode PSOAM waves should be equal to the $\Delta l_e$ between multiplexed MGs. For the proposed MG-MIMO, the multiplexed MGs are $\rm MG \left\lbrace-4,-3,-2,-1\right\rbrace$ and $\rm MG \left\lbrace 1,2,3,4\right\rbrace$ with $\Delta l_e=5$. As for the PSOAM-MIMO, the multiplexed single mode PSOAM wave are $l=+2$ and $l=-3$. In addition, when the MG antenna is used for radiating single mode PSOAM wave, the RF signals carrying data streams can be directly fed into the ports of corresponding PSOAM modes without going through the feeding network, meanwhile, the remaining ports of the MG antenna are terminated with several matching loads. By contrast, for the proposed MG-MIMO, a higher insertion-loss is going to be introduced actually.
\begin{figure}[t]
%	\vspace*{-3mm}
	\centering
	\includegraphics[width=3.0in]{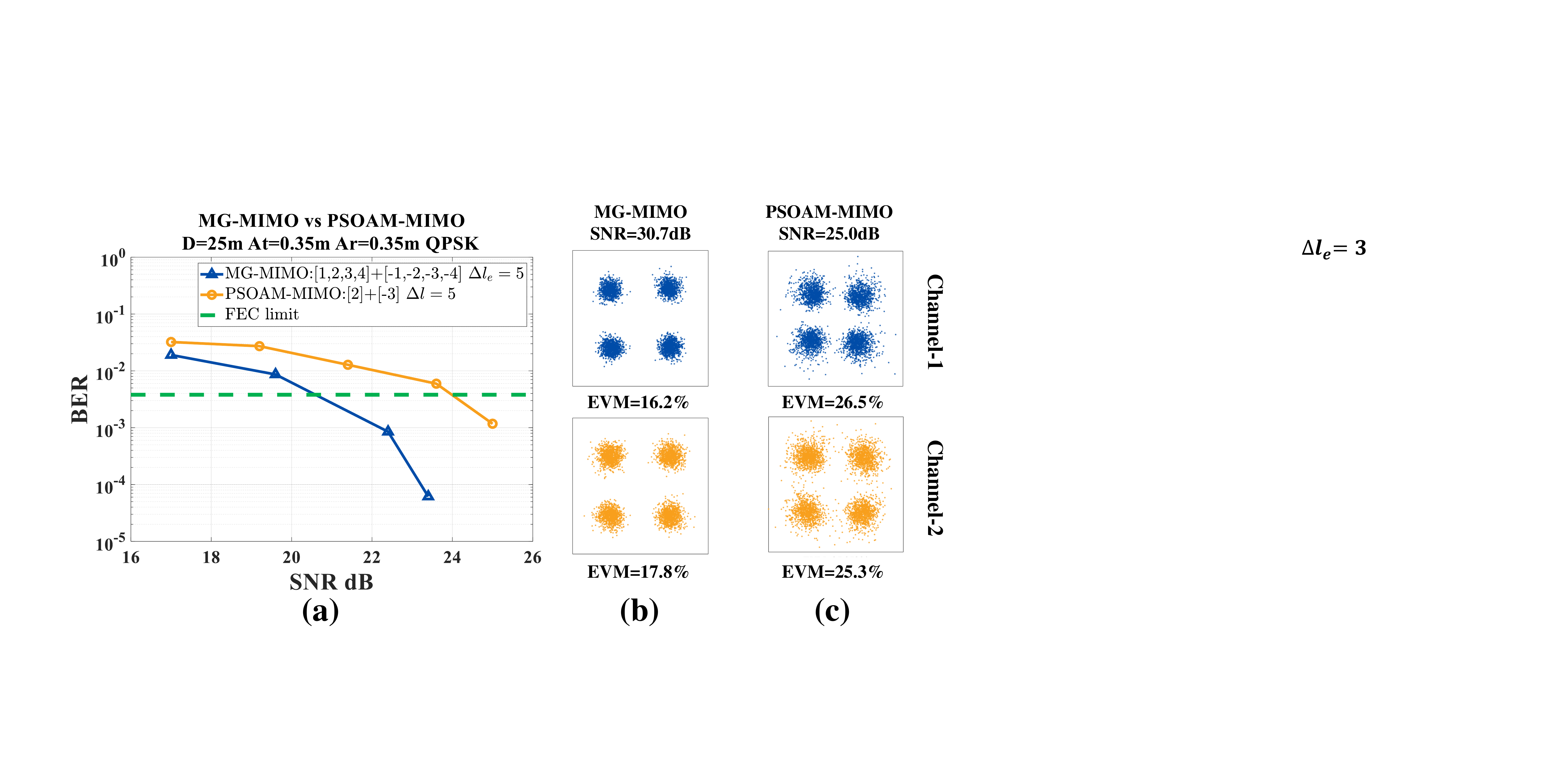}
	\caption{(a) The BER performance of MG-MIMO and PSOAM-MIMO for the QPSK transmission. When the power of transmitting IF signal, the LO power and the gain of receiving LNA are respectively set as $-24$ dBm, 10 dBm and 67 dB, the measured receiving SNR for MG-MIMO and PSOAM-MIMO link are 30.7 dB and 25 dB,respectively; (b) the received constellations of the former at the corresponding SNR; (c) the received constellations of the latter at the corresponding SNR.}
	\label{fig11}
\end{figure}
\begin{figure}[t]
	%	\vspace*{-3mm}
	\centering
	\includegraphics[width=3.2in]{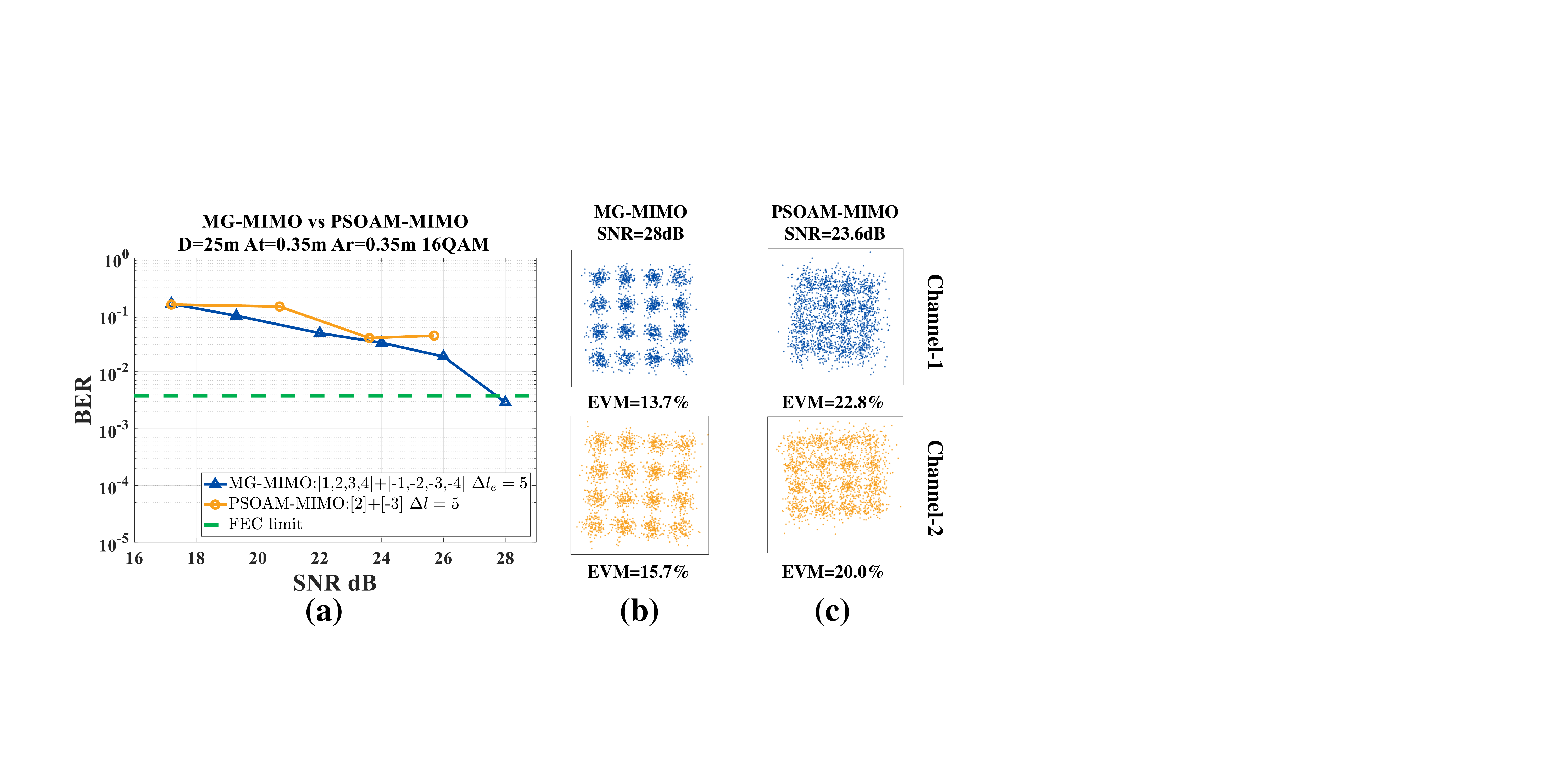}
	\caption{(a) The BER performance of MG-MIMO and PSOAM-MIMO for the 16-QAM transmission. When the power of transmitting IF signal, the LO power and the gain of receiving LNA are respectively set as $-27$ dBm, 10 dBm and 67 dB, the measured receiving SNR for MG-MIMO and PSOAM-MIMO link are 28 dB and 23.6 dB,respectively; (b) the received constellations of the former at the corresponding SNR; (c) the received constellations of the latter at the corresponding SNR.}
	\label{fig12}
\end{figure}

Fig. \ref{fig11}(a) shows the measured raw BER curves of MG-MIMO and PSOAM-MIMO system for the QPSK transmission. In contrast to the PSOAM-MIMO, the MG-MIMO shows a better BER capability in a remote wireless communication scenario. At a receiving SNR of 22.4 dB, MG-MIMO can reach the FEC limit, and PSOAM-MIMO needs a receiving SNR of 25 dB to reach the same limit. For both the systems, when the power of transmitting IF signal, the LO power and the gain of receiving LNA are respectively set as $-24$ dBm, 10 dBm and 67 dB, the attainable SNR of MG-MIMO link is 30.7 dB, and this value of PSOAM-MIMO is only 25 dB, which profits from that the MG constructed by 4 PSOAM waves has a azimuthal beam gain of about 6 dB\cite{zhu2020compact} relative to the single mode PSOAM wave. As shown in Fig. \ref{fig11}(b), at the SNR of 30.7 dB, the EVM of two channels for MG-MIMO are $16.2\%$ and $17.8\%$. Fig. \ref{fig11}(c) shows the EVM performance of two channels for PSOAM-MIMO, which are $26.5\%$ and $25.3\%$ at the SNR of 25 dB.

Likewise, Fig. \ref{fig12}(a) shows the measured raw BER curves of MG-MIMO and PSOAM-MIMO system for the 16-QAM transmission. MG-MIMO can reach the FEC limit at the receiving SNR of 28 dB. In terms of the existing experimental architecture, PSOAM-MIMO is no longer available for reaching the FEC limit owing to the lower SNR. When the power of transmitting IF signal, the LO power and the gain of receiving LNA are respectively set as $-27$ dBm, 10 dBm and 67 dB, the attainable SNR of MG-MIMO link is 28 dB, and this value of PSOAM-MIMO is only 23.6 dB. Here, owing to the higher peak-to-average ratio of 16-QAM, in order to avoid the non-linear distortion\cite{1634783} caused by the RF amplifier, the power of transmitting IF signal is appropriately reduced to $-27$ dBm.  As shown in Fig. \ref{fig12}(b), at the certain SNR of 28 dB, the EVM of two channels for MG-MIMO are $13.7\%$ and $15.7\%$. 
Meanwhile, as we can see from Fig. \ref{fig11}(c), the EVM performance for PSOAM-MIMO are $22.8\%$ and $20.0\%$ at the certain SNR of 23.6 dB.

%\vspace*{-5mm}
\subsection{Orthogonality between multiplexed MGs under PASR scheme}
\begin{figure}[t]
	%	\vspace*{-3mm}
	\centering
	\includegraphics[width=2.2in]{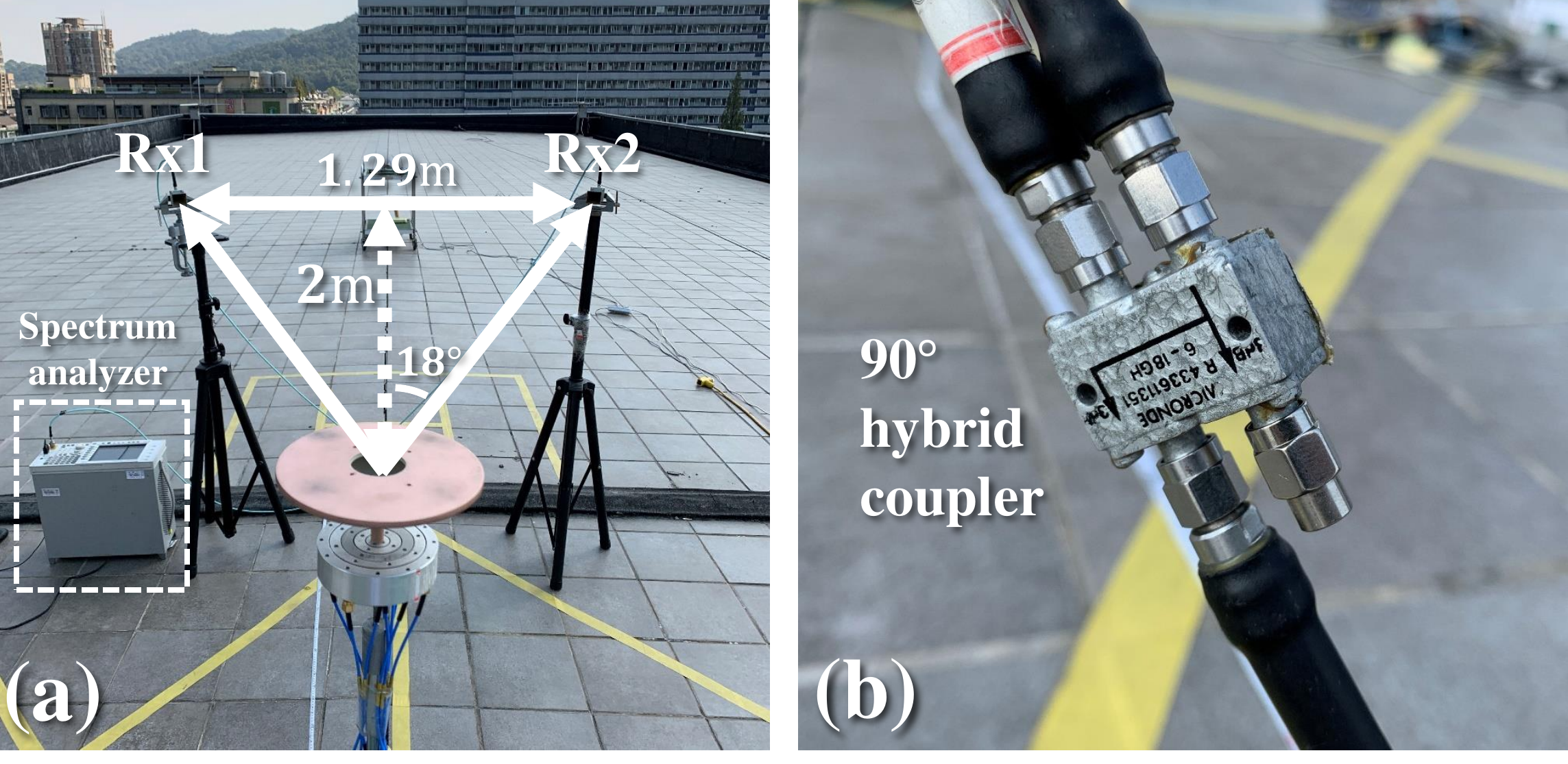}
	\caption{(a) The experimental setup used to evaluate the orthogonality between multiplexed MGs under PASR scheme. (b) The $90^\circ$ hybrid coupler is used to realize the operation of analog phase shifting at the receiving end.}
	\label{fig13}
\end{figure}

According to (\ref{formula4}), if the total number $Q$ and the mode interval $\Delta l$ of multiplexed MGs are same, the generated beams will have the same azimuthal intensity pattern. For the case of $\Delta l=1$, the realized high gain pencil beams present a spiral phase distribution within the mainlobe. Base on the phase information, in virtue of the partial arc sampling receiving (PASR) scheme\cite{zhang2016orbital}, we can utilize the method of analog phase shift to demultiplex data streams orthogonally with a low complexity. In order to use PASR scheme, the number of multiplexed MGs should be equal to the number of receiving antennas and the receiving antennas need to be placed uniformly at the equal-gain position within MG's mainlobe. For $2\pi/\delta$ azimuthal arc and $N_r$ receiving antennas, the angle difference between neighbouring antennas meets $2\pi/(\delta N_r)$. If $\Delta l_e^{n_1 n_2}=l_e^{n_1}-l_e^{n_2}$ is equal to $k'\delta$, where $k'$ is a non-zero integer, the multiplexed $\rm MG^{n1}$ and $\rm MG^{n2}$ can be demultiplexed orthogonally\cite{2020Performance}:
%{\setlength\abovedisplayskip{6pt}
%	\setlength\belowdisplayskip{2pt}
\begin{equation}
	\begin{split}
	\label{formula7}
		\sum_{n_r=1}^{N_r}BP_{\rm MG^{n1}}(\frac{2\pi n_r}{\delta N_r})BP_{\rm MG^{n2}}^{*}(\frac{2\pi n_r}{\delta N_r})\\
		=\gamma \sum_{n_r=1}^{N_r}e^{-j\cdot\frac{2\pi n_r}{N_r}\cdot\frac{l_e^{n_1}-l_e^{n_2}}{\delta}}=0
	\end{split}
\end{equation}
where $\gamma$ contains all constant factor. From (\ref{formula7}), the angle difference between adjacent receiving antennas is limited by the $\Delta l_e$ between multiplexed MGs, in other words, PASR scheme needs a specific receiving aperture $A_r$. For the case of two MGs multiplexing, the multiplexed MGs are $\rm MG \left\lbrace-4,-3,-2,-1\right\rbrace$ and $\rm MG \left\lbrace 1,2,3,4\right\rbrace$ with $\Delta l_e=5$, the specific $A_r$\cite{2020Performance} can be described as 
\begin{equation}
	\label{formula8}
	A_r=2D\cdot\tan(\frac{\varphi_s}{2})=2D\cdot \tan(\frac{\pi}{2\Delta l_e})
\end{equation}
where $\varphi_s$ is the angle difference in azimuthal direction between neighbouring receiving antennas. Fig. \ref{fig13}(a) shows the corresponding experimental setup, when $D$ is set as 2 m, the angle difference between neighbouring antennas can be calculated according to (\ref{formula8}), its value is equal to $36^\circ$, so $A_r$ should be set as 1.29 m. 
\begin{figure}[t]
	%	\vspace*{-3mm}
	\centering
	\includegraphics[width=1.8in]{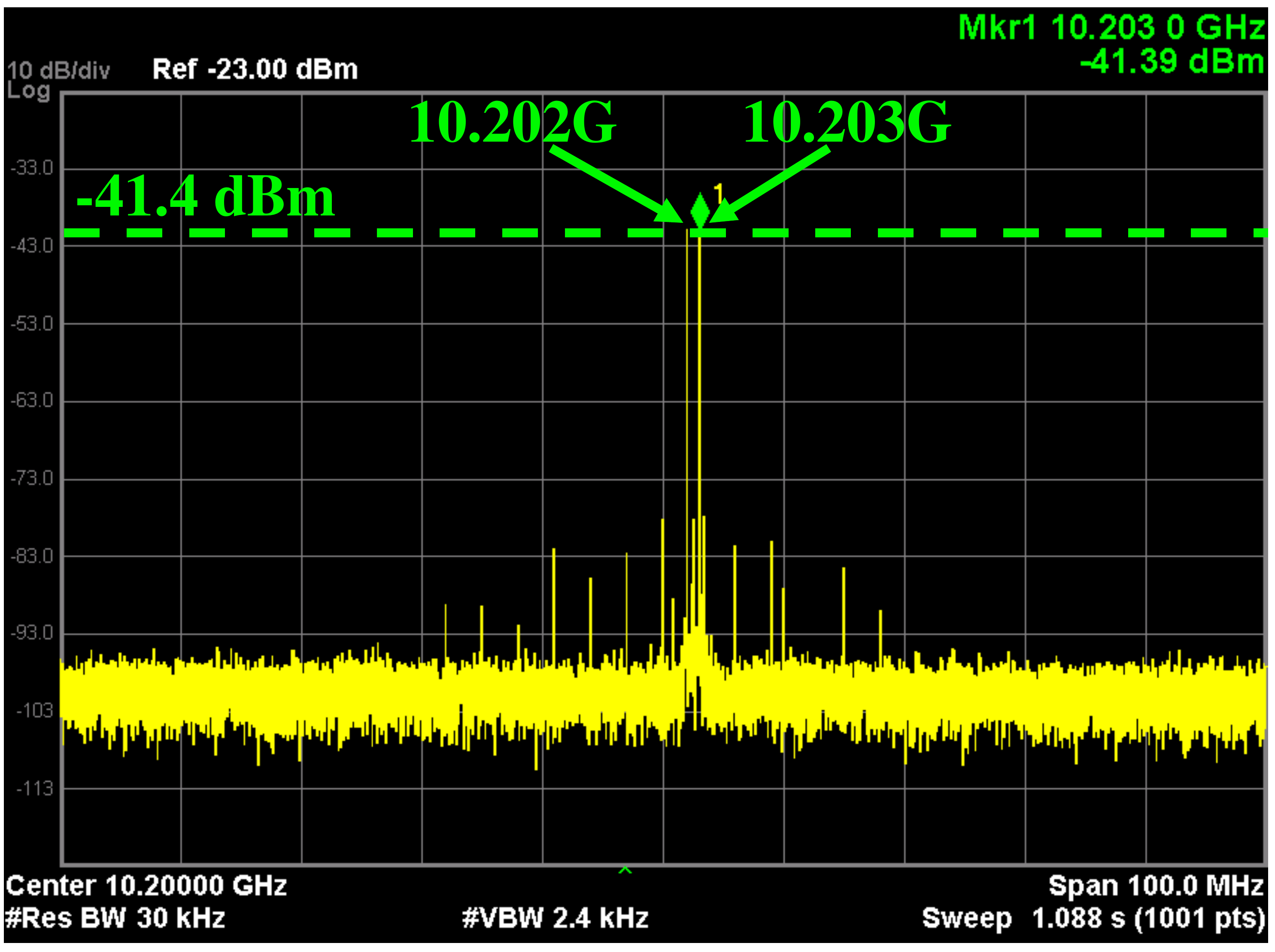}
	\caption{Without the operation of analog phase shift, the received signals at the Rx1, the result is consistent at the Rx2. The key setting parameters of spectrum analyzer: the center frequency (CF) is set as 10.2 GHz; the span is set as 100 MHz; the resolution bandwidth (RBW) is set as 30 KHz; the video bandwidth (VBW) is set as 2.4 KHz.}
	\label{fig14}
\end{figure}

In this experiment, the orthogonality can be characterized by measuring the crosstalk (CT) \cite{hui2015multiplexed,zhang2016mode} between two MGs. At the transmitting end, a 10.202 GHz CW signal with the power of 10 dBm is fed to the $\rm MG \left\lbrace 1,2,3,4\right\rbrace$, the other 10.203 GHz CW signal with the same power is fed to the $\rm MG \left\lbrace-4,-3,-2,-1\right\rbrace$. At the receiving end, utilizing the PASR scheme, a $2\times 2$ power transfer matrix could be obtained by measuring the power of both received signals with the help of spectrum analyzer (Agilent N9020A). Further, the CT between multiplexed MGs can be calculated by adding the received power of all other MGs (unexpected) and dividing it by the received power of the corresponding MG (expected). For instance, to calculate the crosstalk of the $\rm MG \left\lbrace 1,2,3,4\right\rbrace$, the received power $P_{unexpected}$ of $\rm MG \left\lbrace-4,-3,-2,-1\right\rbrace$ should be measured and then divided by the received power $P_{expected}$ of $\rm MG \left\lbrace 1,2,3,4\right\rbrace$, it can be given by 
\begin{equation}
	\label{formula9}
	CT=\frac{P_{unexpected}}{P_{expected}}
\end{equation}

According to \cite{zhang2016orbital}, the phase shift is set as $0^\circ$ at the Rx1 and $90^\circ$ at the Rx2 to demultiplex $\rm MG \left\lbrace 1,2,3,4\right\rbrace$. Meanwhile, the phase shift is set as $90^\circ$ at the Rx1 and $0^\circ$ at the Rx2 to demultiplex the other one. Here, as shown in Fig. \ref{fig13}(b), the $90^\circ$ hybrid coupler is used to achieve the corresponding analog phase shift. Certainly, the same functionality can be realized by combining the power combiner and adjustable phase shifter. 
\begin{figure}[t]
	%	\vspace*{-3mm}
	\centering
	\includegraphics[width=3.4in]{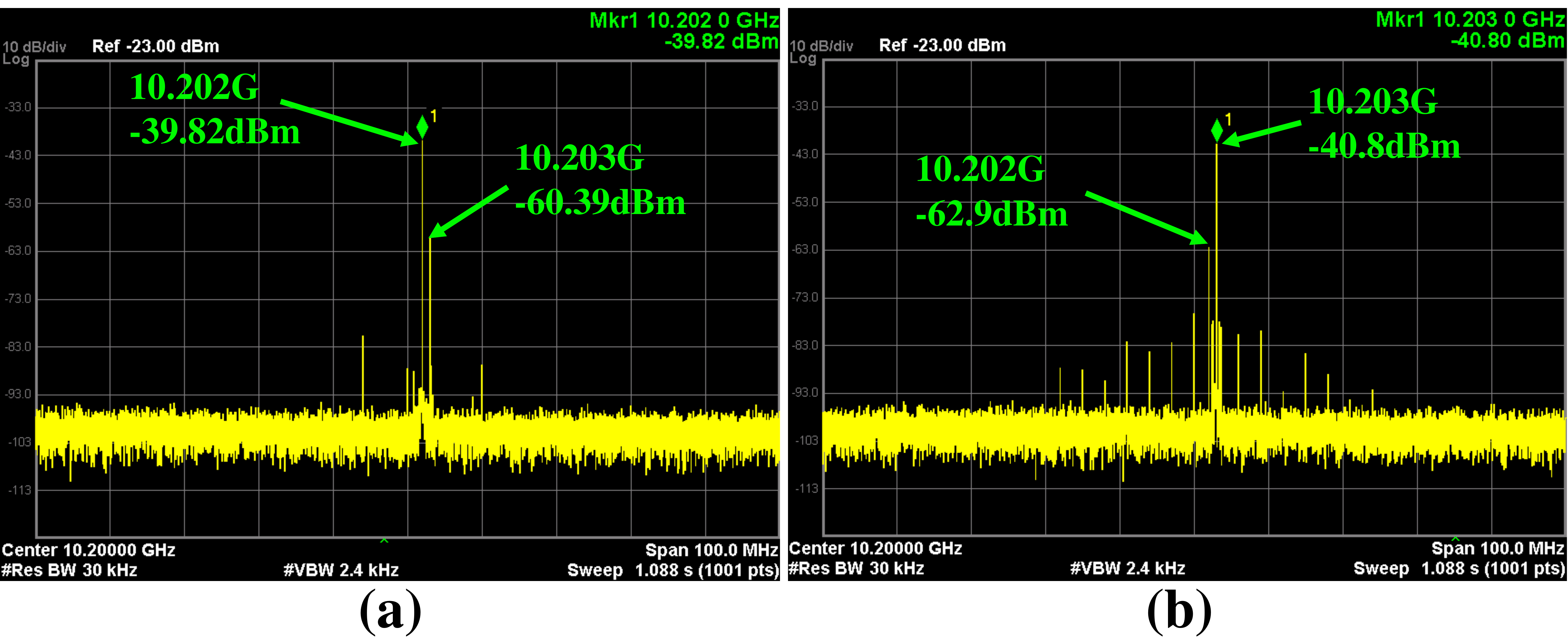}
	\caption{(a) Using the PASR scheme to demultiplex the $\rm MG \left\lbrace 1,2,3,4\right\rbrace$ carrying 10.202 GHz CW signal. (b) Using the PASR scheme to demultiplex the $\rm MG \left\lbrace -4,-3,-2,-1\right\rbrace$ carrying 10.203 GHz CW signal. The  setting parameters of spectrum analyzer remain the same.}
	\label{fig15}
\end{figure}
\begin{table}[t]
	%	\vspace*{-1.0mm}
	\begin{center}
		\caption{The measured power transfer matrix under PASR scheme}
		\begin{tabular}{c|c c}
			\toprule[2pt]
			\textbf{Received power (dBm)}&[Rx1-$0^\circ$,Rx2-$90^\circ$]&[Rx1-$90^\circ$,Rx2-$0^\circ$]\\
			\hline
			Tx: 10.202 GHz&-39.82&-62.9\\ 
			Tx: 10.203 GHz&-60.39&-40.8\\ 
			\hline
			CT (dB)&\textbf{-20.57}&\textbf{-22.1}\\
			\bottomrule[2pt]
		\end{tabular}\\
%		\vspace*{+1.0mm}
%		\footnotesize{$^a$$f_c$ is the carrier frequency; $^b$$A_t$ is the transmitting aperture; and $^c$$A_t$ is the receiving aperture.}\\
		\label{tab2}
	\end{center}
\end{table}
If the analog phase shift based on PASR scheme is not operated, as shown in Fig. \ref{fig14}, the power level of two CW signals are all maintained around -41.4 dBm. In virtue of PASR scheme, Fig. \ref{fig15}(a) shows the demultiplexing of the 
$\rm MG \left\lbrace 1,2,3,4\right\rbrace$ carrying 10.202 GHz CW signal, its measured intensity is -39.82 dBm, while the intensity of 10.203 GHz CW is suppressed to -60.39 dBm. In addition, Fig. \ref{fig15}(b) shows the demultiplexing of the $\rm MG \left\lbrace-4,-3,-2,-1\right\rbrace$ carrying 10.203 GHz CW signal, its measured intensity is -40.8 dBm, conversely, the intensity of 10.202 GHz CW is suppressed to -62.9 dBm. Table \ref{tab2} shows the measured power transfer matrix, the CT of both the CW signals are below -20 dB. Therefore, it has been proved that the isolation under PASR scheme between two MGs is at least 20 dB and multiplexed MGs have orthogonality to some degree. 

Unfortunately, for a long distance transmission and such a low $\Delta l_e$, the PASR scheme needs a big and exaggerated $A_r$. For example, for above-mentioned two multiplexed MGs, if the communication distance $D$ is extended to 10 m, the corresponding $A_r$ under PASR scheme will be 6.5 m, which is clearly an unacceptable $A_r$ for practical application. So, there is a trade-off between the receiving complexity and the engineering feasibility. Assuming that two MGs with a $\Delta l_e$ of 80 can be generated, for the same $D$ of 10 m, the corresponding $A_r$ under PASR scheme could be reduced to 0.4 m. Therefore, it's necessary to generate a MG with higher $l_e$, in \cite{2020Direct}, partial arc transmitting (PAT) scheme provides a feasible method to achieve this goal.

\subsection{Comparison between conventional MIMO and MG-MIMO}

\begin{figure}[t]
	%	\vspace*{-3mm}
	\centering
	\includegraphics[width=3.4in]{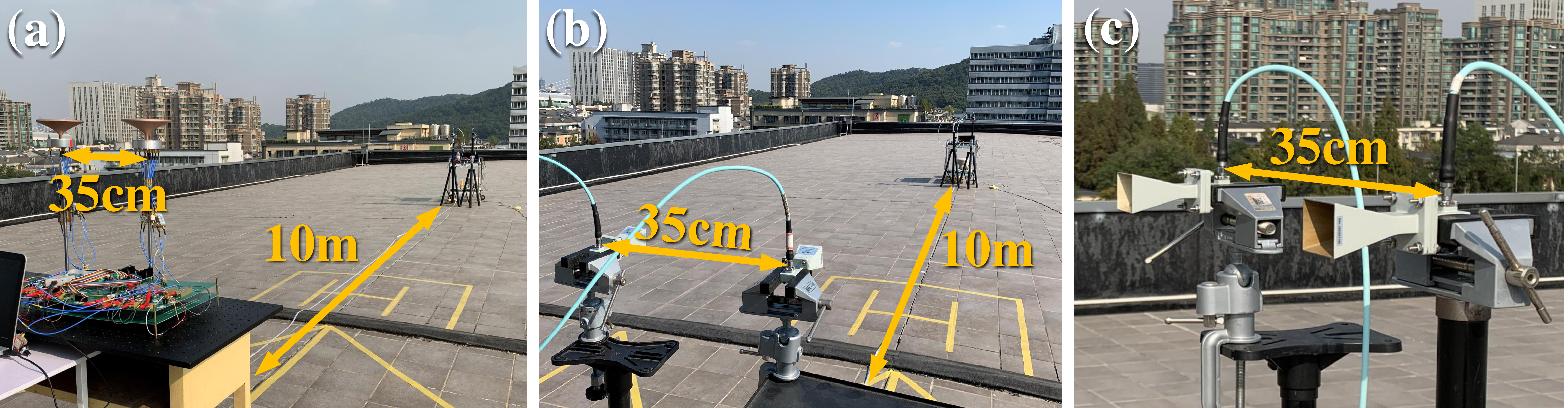}
	\caption{The comparison between the conventional MIMO and the proposed MG-MIMO. (a) The experimental setup of MG-MIMO link. (b) The experimental setup of MIMO link. (c) The receiving end for both the links.}
	\label{fig16}
\end{figure}

In this sub-section, the comparison of communication performance for the conventional MIMO link and the proposed MG-MIMO link have been demonstrated and analyzed under the same channel condition. Fig. \ref{fig16}(a) and Fig. \ref{fig16}(b) show the experimental setup of the proposed MIMO and the conventional MIMO, respectively. Particularly, Fig. \ref{fig16}(c) shows the receiving end setup for both the links. The transmitting end for both links adopt the ULA arraying way, different from the MG-MIMO, the array element at the transmitting end in MIMO adopts the horn antenna, where the wavefront of the beam generated by the horn can be regarded as an
equi-phase plane. Therefore, in conventional MIMO, the sub-channel difference is only caused by the propagation path difference, in other words, it depends on the aperture of the transceiving end and the communication distance. For the receiving end of both links, two horn antennas are still placed at the symmetry positions along the propagation axis. In MG-MIMO link, a pair of MGs used for this contrast experiment are $\rm MG \left\lbrace 1,2,3,4\right\rbrace$ and $\rm MG\left\lbrace-4,-3,-2,-1\right\rbrace$ with $\Delta l_e=5$. Other significant experimental parameters have been given in Table \ref{tab3}.
\begin{table}[h]
	%	\vspace*{-1.0mm}
	\begin{center}
		\caption{Experimental Parameters for MIMO and MG-MIMO}
		\begin{tabular}{c|c c}
			\toprule[2pt]
			\textbf{Parameters}&\textbf{MIMO}&\textbf{MG-MIMO}\\
			\hline
			$A_t$ (m)&0.35&0.35 \\ 
			$A_r$ (m)&0.35&0.35 \\ 
			$D$ (m)&10&10\\ 
			\hline
			Modulation&\multicolumn{2}{|c}{QPSK, 16-QAM, 64-QAM}\\
			Rayleigh distance (m)&\multicolumn{2}{|c}{$2A_r^2/\lambda\approx8.3$} \\
			\bottomrule[2pt]
		\end{tabular}
		\label{tab3}
	\end{center}
\end{table}  
\begin{figure}[h]
	%	\vspace*{-3mm}
	\centering
	\includegraphics[width=3.4in]{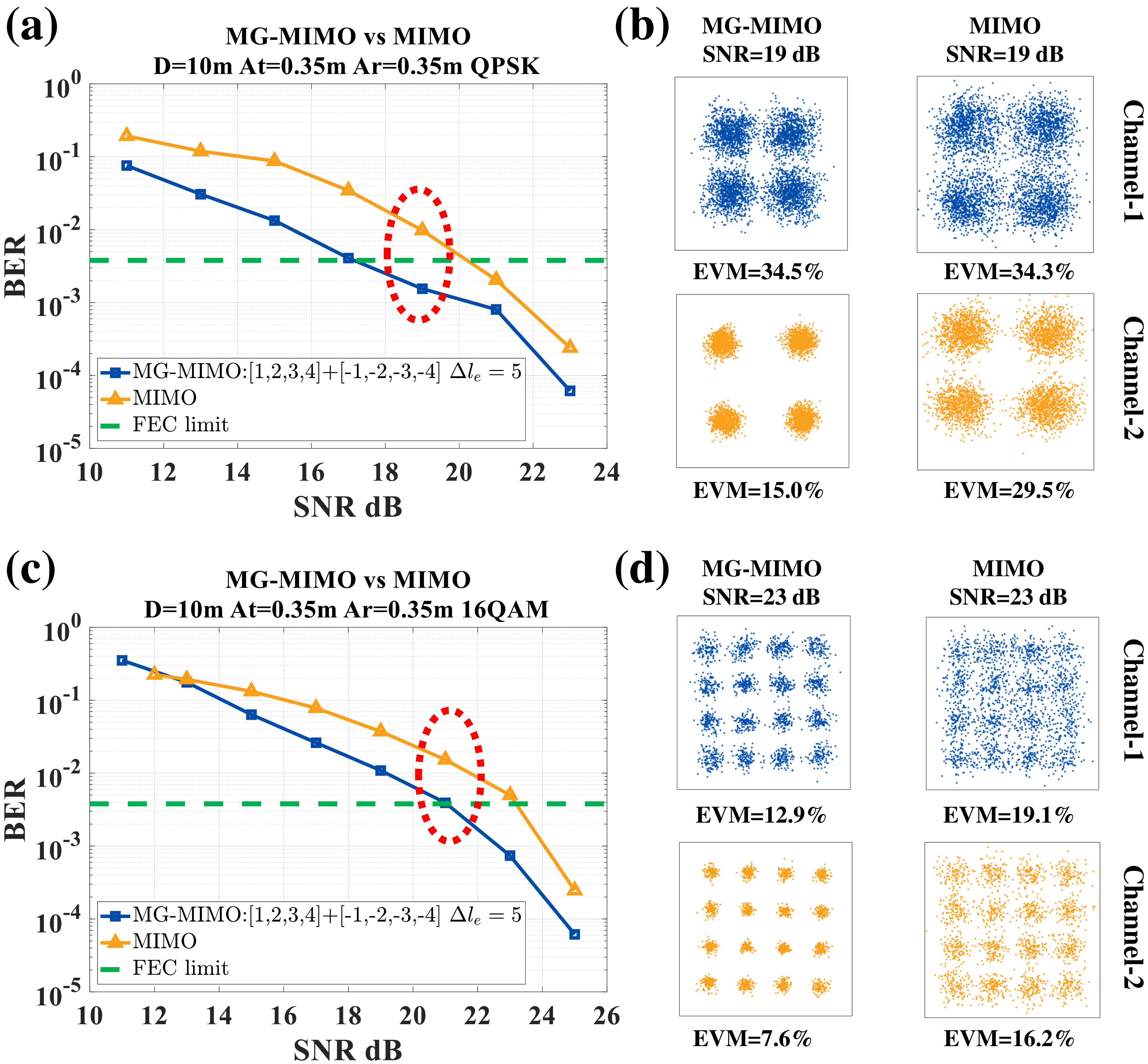}
	\caption{(a) The BER performance of MG-MIMO and MIMO for the QPSK transmission; and (b) the corresponding received constellations at a receiving SNR of 19 dB. (c) The BER performance of MG-MIMO and MIMO for the 16-QAM transmission; and (d) the corresponding received constellations at a receiving SNR of 23 dB.}
	\label{fig17}
\end{figure}

In order to compare the communication performance of MG-MIMO link and MIMO link, the measured raw BER curves of the QPSK transmission and the 16-QAM transmission are shown in Fig. \ref{fig17}(a) and Fig. \ref{fig17}(c), respectively. Besides, Fig. \ref{fig17}(b) shows the typical received constellations with the EVM performance of the QPSK transmission at a same receiving SNR of 19 dB for both the links. Fig. \ref{fig17}(d) shows the corresponding results of the 16-QAM transmission at a same receiving SNR of 23 dB. With the same experimental setup given by Table \ref{tab3}, for both two modulation formats, the MG-MIMO link shows a better BER performance than the MIMO link at the same receiving SNR. Such the improvement comes from the phase control of beam's wavefront manipulated by MG technique, i.e. MG antennas applied for arraying at the transmitting end possess the function of reducing the sub-channel correlation, which means that the MG-MIMO system has the potential of long-distance wireless transmission. On the point of reducing the sub-channel correlation, the PSOAM-MIMO can also achieve the same effect as the MG-MIMO\cite{zhang2017experimental}. However, to our knowledge, PSOAM-MIMO system has two obvious flaws: one is the deterioration of receiving SNR caused by the waste of energy for the same certain transmitting power; the other is that single mode PSOAM antenna is lacking in the flexible control\cite{zheng2017realization} of wavefront's phase distribution. In \cite{zhang2017experimental}, despite showing a better BER performance compared with conventional MIMO, but $A_t$ and $A_r$ of PSOAM-MIMO system are all 0.525 m, the corresponding Rayleigh distance for the operating frequency of 10 GHz is about 18.4 m, its actual transmission distance $D$ is merely 7 m less than half the Rayleigh distance.

For the QPSK transmission, the MG-MIMO link can reach the FEC limit at a lower SNR of about 19 dB, and the conventional MIMO link needs a 21 dB SNR to reach the same limit. Besides, at the same receiving SNR of 19 dB, the EVM of two channels for MG-MIMO are $34.5\%$ and $15.0\%$, and the EVM performance for MIMO are $34.3\%$ and $29.5\%$. Furthermore, as far as the test results are concerned, for the 16-QAM transmission, the MG-MIMO link can reach the FEC limit at a lower SNR of about 23 dB, and the conventional MIMO link requires a 25 dB SNR to reach the same limit. At the same receiving SNR of 23 dB, the EVM of two channels for MG-MIMO are $12.9\%$ and $7.6\%$, and the EVM performance for MIMO are $19.1\%$ and $16.2\%$. 

\begin{figure}[t]
	%	\vspace*{-3mm}
	\centering
	\includegraphics[width=3.4in]{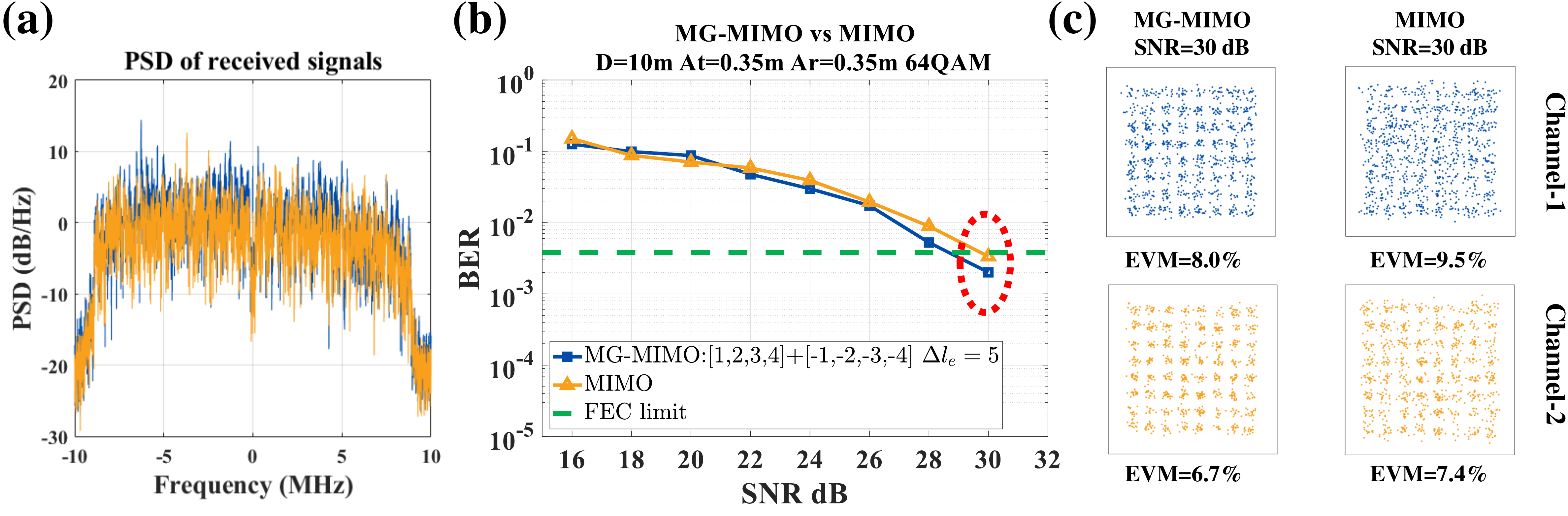}
	\caption{(a) The typical PSD of the received signals at a receiving SNR of 30 dB for the 64-QAM transmission. (b) The measured BER performance of MG-MIMO and MIMO for the 64-QAM transmission. (c) The corresponding received constellations at a receiving SNR of 30 dB.}
	\label{fig18}
\end{figure}

As for the higher order modulation format, such as the 64-QAM, Fig. \ref{fig18}(a) illustrates the typical received PSD with a BW of 20 MHz at a SNR of 30 dB. Distinguished from the PSD for lower order modulation formats shown in Fig. \ref{fig8}, the OFDM signal with the 64-QAM modulation format has a higher peak-to-average ratio significantly. The measured raw BER curves of the 64-QAM transmission are shown in Fig. \ref{fig18}(b). Fig. \ref{fig18}(c) shows the received constellations with the measured EVM at a same receiving SNR of 30 dB for both the links. When the actual $D$ exceeds the Rayleigh distance, the correlation between sub-channels increases dramatically\cite{edfors2011orbital}, the signal multiplexing is becoming more difficult. Compared with the MIMO link, the MG-MIMO link shows a little advantage of BER performance only in the receiving SNR range from 26 dB to 30 dB. In terms of the experimental results, both the links reach the FEC limit at a receiving SNR of 30 dB. At this SNR, the EVM of two channels for MIMO link are $9.5\%$ and $7.4\%$, because it's merely just over the FEC limit, the signal clusters of channel-1 are not converged very well. Whereas, the MG-MIMO link shows a clearer signal cluster distribution, its corresponding EVM performance are $8.0\%$ and $6.7\%$. An OFDM symbol consists of 56 sub-carriers, of which 52 sub-carriers are used for the data transmission in this experiment. Accordingly, like most of practical OFDM systems calculated\cite{zhang2016mode}, the attainable spectrum efficiency of the 64-QAM transmission for the MG-MIMO link would be 11.1 bit /s/Hz (6 bits per symbol $\times$ 2 data streams $\times$ $\frac{52}{56}$).

Owing to the perturbation and the fading caused by time-varying characteristic of wireless channel\cite{goldsmith2005wireless}, even if the test channel is a LoS scenario, the robustness of the proposed MG-MIMO link should be evaluated to analyze the persistence of the system’s characteristics. Here, for three modulation formats, we go through the measurement of multiple data frames transmission and then count the number of correctly transmitted frames among them. The cyclic redundancy check \cite{castagnoli1993optimization} algorithm is applied for the receiving end to determine whether a data frame is transmitted correctly. When the value of CRC is equal 1, it's verified that the data frame transmission is correct. Here, being subject to the RAM limitation of upper-computer, the total number of data frames is set as 250. 

\begin{figure}[t]
	%	\vspace*{-6mm}
	\centering
	\includegraphics[width=3.4in]{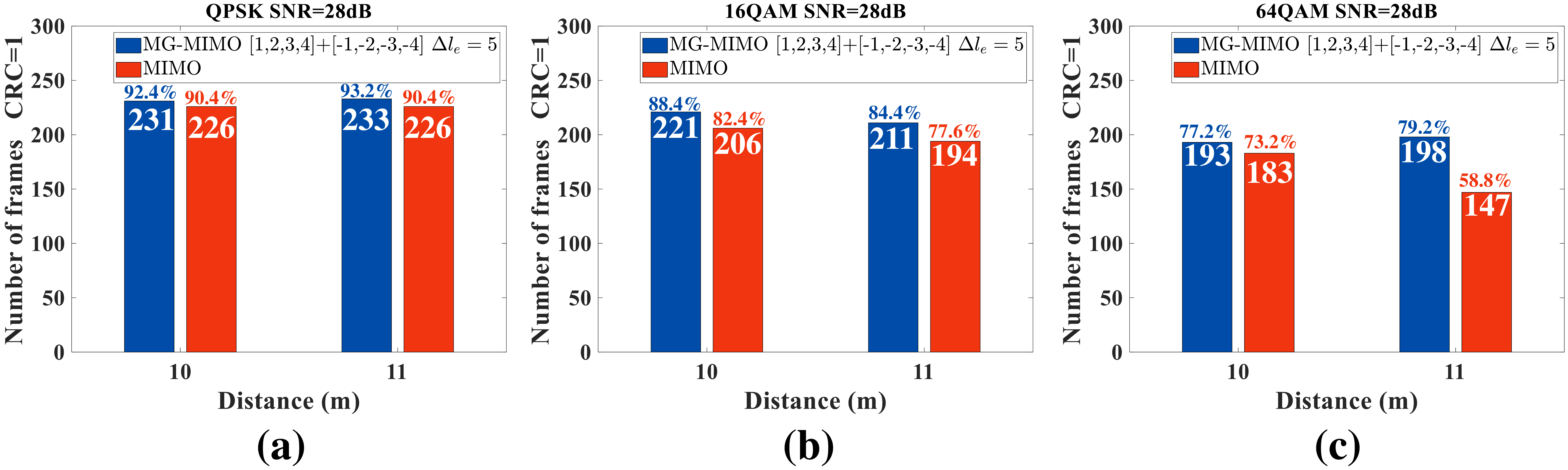}
	\caption{The results of system robustness testing at a receiving SNR of 28 dB. Both the measuring distances are over the Rayleigh distance.}
	\label{fig19}
\end{figure}

Fig. \ref{fig19} shows the robustness testing results for both links at a certain receiving SNR of 28 dB. For each $D$ and each modulation format, when the SNR is fixed, the robustness of MG-MIMO is better than that of MIMO. For the low order modulation such as QPSK, as $D$ increases, the transmission success-rate $\eta_s$ of both links are maintained at a stable level, when D is 10 m, the $\eta_s$ of MG-MIMO and MIMO are $92.4\%$ and $90.4\%$, respectively, when D is 11 m, this value for both links keep almost unchanged. For the 16-QAM transmission, the $\eta_s$ of both links are all decreased relative to the QPSK transmission, when D is 10 m, the $\eta_s$ of MG-MIMO and MIMO drop to $88.4\%$ and $82.4\%$, respectively, when D is 11 m, this value for both links drop to $84.4\%$ and $77.6\%$. When it comes to the 64-QAM transmission, although the $\eta_s$ of MG-MIMO has also gone down relative to the 16-QAM transmission, yet its values maintain the level of $77.2\%$ and $79.2\%$ for a $D$ of 10 m and 11 m, respectively. Whereas, as $D$ increases, the $\eta_s$ performance of conventional MIMO deteriorates sharply. When $D$ is 10 m, the $\eta_s$ of MIMO can be more than 70 percent, however, when $D$ increases to 11 m, its  $\eta_s$ goes down to $58.8\%$, which means that nearly half of the data frames are not successfully transmitted.

%\vspace*{8mm}
\section{Demonstration of Long-distance transmission experiment}

\begin{figure}[t]
	%	\vspace*{-6mm}
	\centering
	\includegraphics[width=3in]{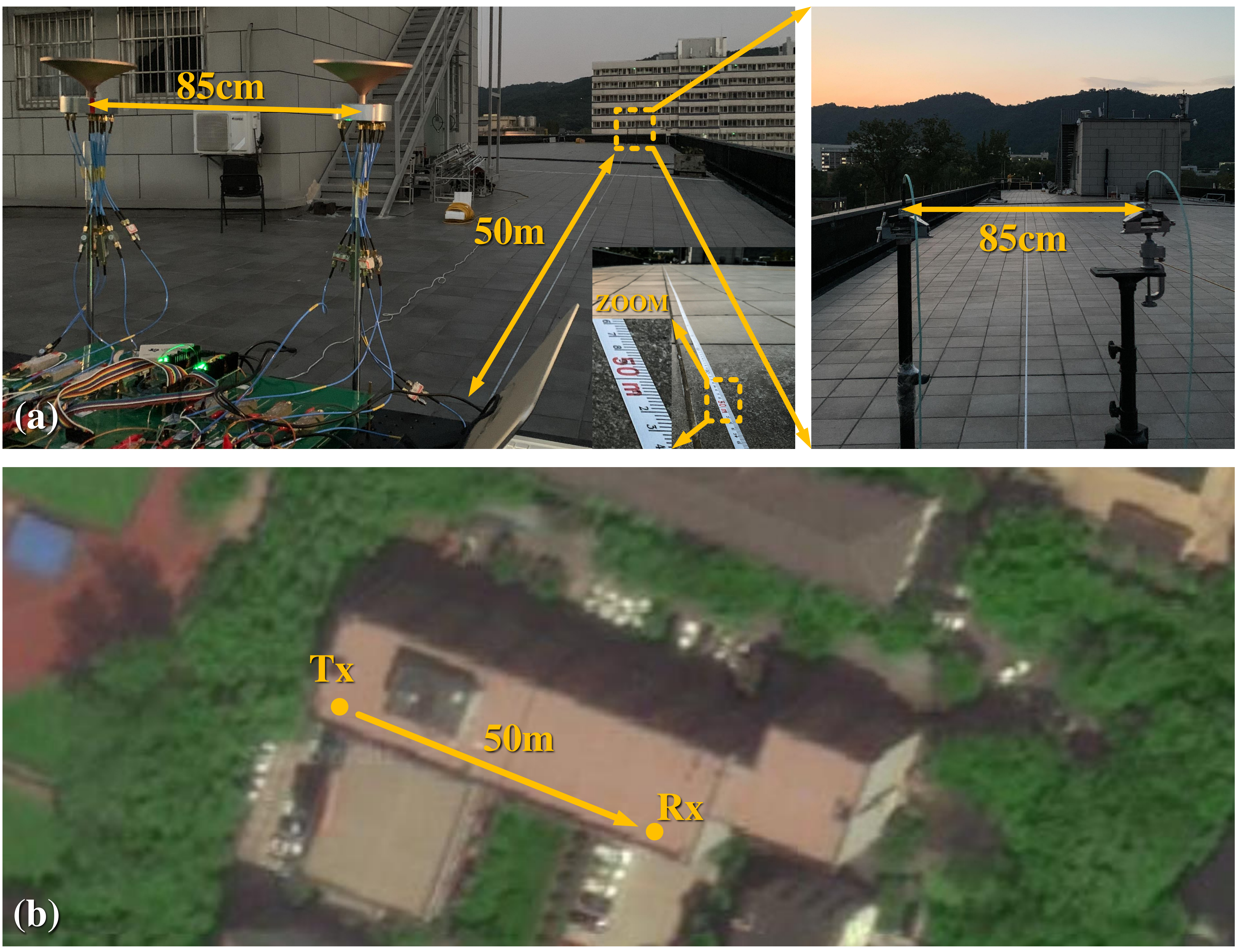}
	\caption{(a) The experimental setup of the long-distance PSOAM MG-MIMO communication link. The aperture of the transmitting end $A_t$ and the rseceiving end $A_r$ are all 0.85 m. The operating frequency is 10.2 GHz, the corresponding Rayleigh distance ($2A_r^2/\lambda$) is about 49 m. The actual $D$ is 50 m. (b) The satellite imagery of the experiment site in Hangzhou. The experimental site is selected on a rooftop in order to create a better LoS scene, which is located in Yuquan Campus of Zhejiang University.} 
	\label{fig20}
\end{figure}

In this section, we attempt to perform a long-distance MG-MIMO communication link. Both the transmitting and receiving end are going to adopt the ULA arraying way, $A_t$ and $A_r$ are all set as 0.85 m. For the operating frequency of 10.2 GHz, the corresponding Rayleigh distance ($2A_r^2/\lambda$) can be calculated as about 49 m\cite{balanis2016antenna}. In this experiment, the multiplexed PSOAM MGs are $\rm MG \left\lbrace-4,-3,-2,-1\right\rbrace$ and $\rm MG \left\lbrace 1,2,3,4\right\rbrace$ with $\Delta l_e=5$. The modulation formats are QPSK and 16-QAM. The out-door experimental setup of this long-distance transmission has been shown in Fig. \ref{fig20}.

\begin{figure}[t]
	%	\vspace*{-6.5mm}
	\centering
	\includegraphics[width=3.2in]{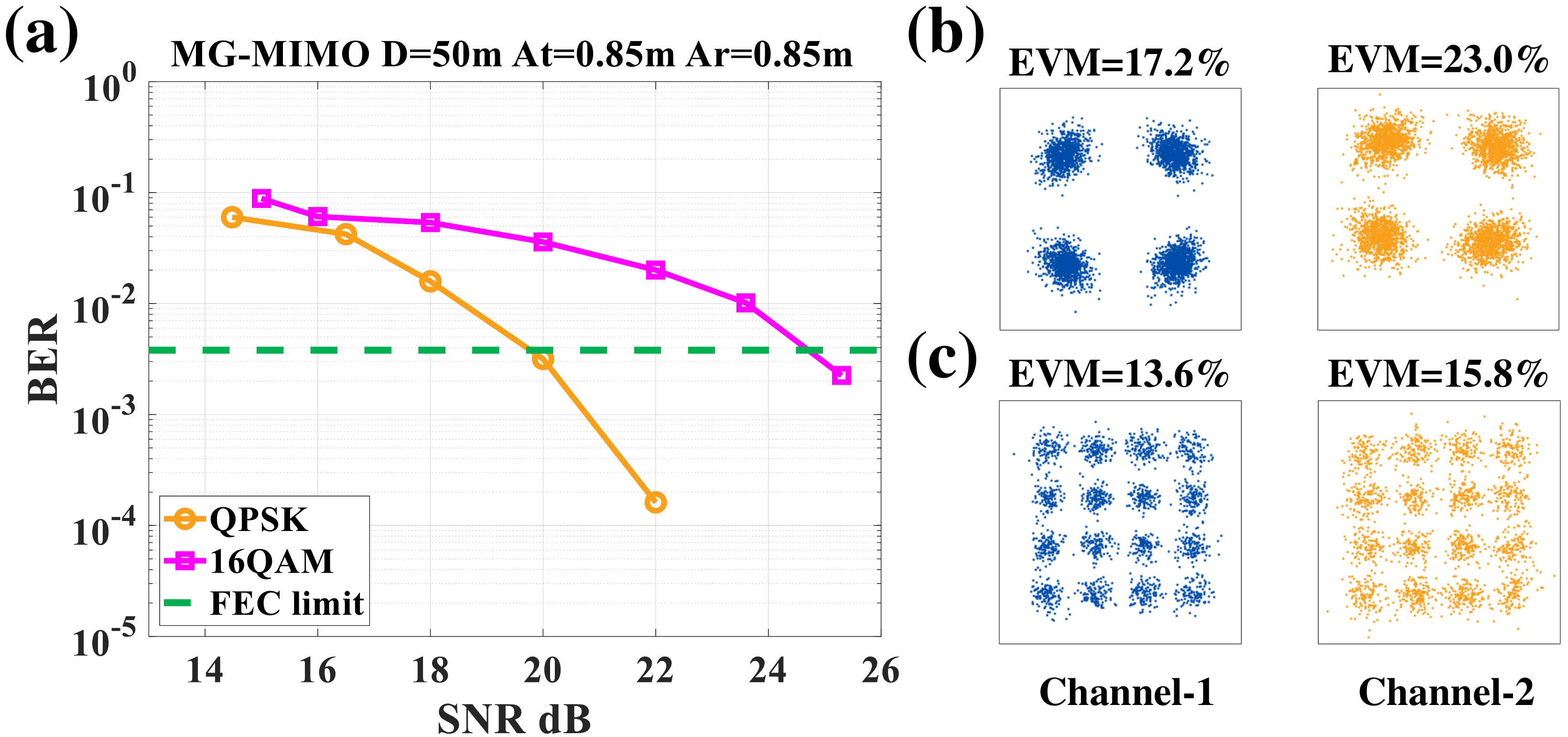}
	\caption{(a) The measured raw BER curves of MG-MIMO with a $D$ of 50 m for the QPSK and the 16-QAM transmission. The typical received constellations with measured EVM performance (b) for the QPSK transmission at a receiving SNR of 20 dB and (c) for the 16-QAM transmission at a receiving SNR of 25.3 dB.}  
	\label{fig21}
\end{figure}

Fig. \ref{fig21}(a) shows the measured raw BER curves of this long-distance MG-MIMO link for the QPSK and the 16-QAM transmission. For the QPSK transmission, at a certain receiving SNR of 20 dB, MG-MIMO can reach the FEC limit. Besides, for the 16-QAM transmission, MG-MIMO needs a receiving SNR of 25.3 dB to reach the same limit. Moreover, Fig. \ref{fig21}(b) shows the typical received constellations with the measured EVM performance of two channels for the QPSK transmission. At a receiving SNR of 20 dB, the EVM of two channels for the QPSK transmission are $17.2\%$ and $23.0\%$. Meanwhile, Fig. \ref{fig21}(c) shows the corresponding received constellations for the 16-QAM transmission, at a receiving SNR of 25.3 dB, the measured EVM of two channels are $13.6\%$ and $15.8\%$, respectively. With a same communication architecture of the preceding proof-of-principle experiments, the attainable spectrum efficiency of the QPSK and the 16-QAM transmission are 3.7 bit /s/Hz (2 bits per symbol $\times$ 2 data streams $\times$ $\frac{52}{56}$) and 7.4 bit /s/Hz (4 bits per symbol $\times$ 2 data streams $\times$ $\frac{52}{56}$).

In addition, it's necessary to set a baseline for comparison.  According to the excellent work in \cite{sasaki2020experimental}, NTT Corporation has successfully performed an OAM multiplexing transmission link at a distance of 100 m. Its operating frequency is 40 GHz, its $A_t$ and $A_r$ are all set as 1.2 m, so the corresponding Rayleigh distance $D_R$ is equal to 384 m. The actual $D$ is about $0.26D_R$. Thanks to their strong engineering capability in industry, they have implemented 15 streams multiplexing, the single-way spectrum efficiency is calculated as 4.4 bits/s/Hz/stream. As for our experimental link, the highest order modulation format that can be transmitted is the 16-QAM, the attainable single-way spectrum efficiency is 3.7 bits/s/Hz/stream. However, for a $D$ of 50 m, the physical aperture $A_r$ of 0.85 m and the operating frequency of 10.2 GHz, our achievable communication distance is about $1.02D_R$ ($D_R=$49 m), which means that MG technique may possess potential in applying to the long-distance point-to-point wireless back-haul communication.

%	\vspace*{-6.5mm}
\section{Conclusion}
In this paper, we have demonstrated a series of proof-of-principle experiments of MG-MIMO system to verify and evaluate the reduction of spatial correlation caused by the MG's vorticity, the SNR improvement caused by the MG's beam directionality gain, the orthogonality between multiplexed MGs under PASR scheme and the better robustness than conventional MIMO when it comes to a communication distance beyond Rayleigh distance. Besides, a tentative long-distance transmission link of MG-MIMO system has been carried out at a communication distance of 50 m with the single-way spectrum efficiency of 3.7 bits/s/Hz/stream. Although the research on the MG based communication technique is still a young-born field, based on the above experimental results, the MG may provide a novel physical layer method to further improve the communication performance in a MIMO system. Especially, it has potential in the long-range wireless back-haul communication link. 

The future investigation on the proposed MG-MIMO system can be focused on the following aspects. As for enhancing the spectrum efficiency, we can increase the number of data streams and adopt the higher order modulation formats to achieve the goal, which requires the updating of baseband communication platform and generating more MGs by optimizing antenna design. The above-mentioned PASR scheme can demultiplex the MGs with a low signal processing complexity, but from a practical point of view, we need to design and fabricate a new antenna \cite{2020Direct} to directly generate a MG with high $l_e$ in place of the superposing method that this paper adopts. This work is currently ongoing.

\ifCLASSOPTIONcaptionsoff
  \newpage
\fi

\bibliographystyle{IEEEtran}
\bibliography{ref}

\end{document}